\documentclass[a4paper,fleqn,usenatbib]{mnras}

\usepackage[T1]{fontenc}
\usepackage{ae,aecompl}

\usepackage{graphicx}	
\usepackage{amsmath}	
\usepackage{amssymb}	
\usepackage{subfig}
\usepackage{journal_names}

\newcommand{\muJy}{\,$\mu\rm{Jy}$}
\newcommand{\mJy}{\,mJy}
\newcommand{\power}[2]{\ensuremath{{#1}{\times}10^{#2}}} 

\title[Dwarf novae are significant radio emitters]{Dwarf nova-type cataclysmic variable stars are significant radio emitters}

\author[D. L. Coppejans et al.]{
Deanne L. Coppejans,$^{1}$\thanks{E-mail: d.coppejans@astro.ru.nl}
Elmar. G. K\"{o}rding,$^{1}$
James C.A. Miller-Jones,$^{2}$\newauthor
Michael P. Rupen,$^{3}$
Gregory R. Sivakoff,$^{4}$
Christian Knigge,$^{5}$
Paul J. Groot,$^{1}$\newauthor
Patrick A. Woudt,$^{6}$
Elizabeth O. Waagen$^{7}$
and Matthew Templeton
\\
$^{1}$Department of Astrophysics/IMAPP, Radboud University, P.O. Box 9010, 6500 GL Nijmegen, The Netherlands\\
$^{2}$International Centre for Radio Astronomy Research, Curtin University, GPO Box U1987, Perth, WA 6845, Australia\\
$^{3}$National Research Council of Canada, Herzberg Astronomy and Astrophysics, Dominion Radio Astrophysical Observatory,\\P.O. Box 248, Penticton, BC V2A 6J9, Canada\\
$^{4}$Department of Physics, University of Alberta, CCIS 4-183, Edmonton, Alberta T6G 2E1, Canada\\
$^{5}$School of Physics and Astronomy, Southampton University, Highfield, Southampton SO17 1BJ, UK\\
$^{6}$Department of Astronomy, University of Cape Town,\\Private Bag X3, 7701 Rondebosch, South Africa\\
$^{7}$American Association of Variable Star Observers, 49 Bay State Road, Cambridge, MA 02138, USA
}

\date{Accepted XXX. Received YYY; in original form ZZZ}

\pubyear{2016}

\begin{document}
\label{firstpage}
\pagerange{\pageref{firstpage}--\pageref{lastpage}}
\maketitle

\begin{abstract}

We present 8--12\,GHz radio light curves of five dwarf nova (DN) type Cataclysmic Variable stars (CVs) in outburst (RX And, U Gem and Z Cam), or superoutburst (SU UMa and YZ Cnc), increasing the number of radio-detected DN by a factor of two. The observed radio emission was variable on time-scales of minutes to days, and we argue that it is likely to be synchrotron emission. This sample shows no correlation between the radio luminosity and optical luminosity, orbital period, CV class, or outburst type; however higher-cadence observations are necessary to test this, as the measured luminosity is dependent on the timing of the observations in these variable objects. The observations show that the previously detected radio emission from SS Cyg is not unique in type, luminosity (in the plateau phase of the outburst), or variability time-scales. Our results prove that DN, as a class, are radio emitters in outburst.
\end{abstract}



\section{Introduction}\label{sec:intro}

Cataclysmic Variable stars (CVs) are binary systems consisting of a white dwarf that accretes matter from a main-sequence secondary star via Roche-lobe overflow \citep[for a review, see][]{Warner1995}. Systems in the dwarf nova (DN) class of CVs show episodic outbursts in which the optical emission of the system brightens by $\sim$2--8 mag over a period of days to weeks. These outbursts develop when a build-up of matter in the accretion disc triggers a thermal-viscous instability and switches the disc to a hot, bright, viscous state \citep[the Disc Instability Model (DIM);][]{Smak1971,Osaki1974,Hoshi1979,Lasota2001}. The interval between subsequent outbursts is weeks to years \citep[e.g.][]{Coppejans2016}.

One of the main questions in CV research is what role mass ejections (in the form of collimated jets or uncollimated winds) play during DN outbursts. Most other classes of accreting objects have been found to launch jets in at least some states. Jets have been found in the symbiotic systems (white dwarfs accreting from a red giant; e.g. \citealt{Sokoloski2008}), neutron star and black hole X-ray Binaries (XRBs; e.g. \citealt{Russell2013}), active galactic nuclei (AGN; e.g. \citealt{King2011}) and gamma-ray bursts (GRBs; e.g. \citealt{Granot2014}). However, until recently it was commonly accepted that CVs do not launch jets. This was used to constrain jet-launching models \citep{Livio1999,Soker2004}.  

As radio emission is the best tracer for jets, it was the historical lack of detections of CVs at radio wavelengths that led to the conclusion that CVs do not launch jets. Although many surveys were conducted prior to 2008, only three non-magnetic CVs\footnote{CVs in which the magnetic field strength of the white dwarf is $B\lesssim10^6\,$G, allowing matter to accrete onto the white dwarf via an accretion disc.} were detected at radio wavelengths. EM Cyg \citep{Benz1989}, SU UMa \citep{Benz1983} and TY Psc \citep{Turner1985} were detected in one set of observations, but were not detected in follow-up observations. Radio surveys of CVs were conducted by e.g. \citet{Benz1983}, \citet{Benz1989}, \citet{Cordova1983}, \citet{Fuerst1986}, \citet{Echevarria1987}, \citet{Nelson1988}, \citet{Turner1985}, and \citet{Benz1996}. The detection rates were higher for the magnetic CVs\footnote{CVs in which the magnetic field strength of the white dwarf is $B>10^6\,$G, which truncates the disc at the Alfven radius.}, but to date only three persistent radio emitters have been identified in the literature, namely AE Aqr \citep[e.g.][]{Bookbinder1987,Bastian1988,Abada-Simon1993,Meintjes2005}, AR UMa \citep{Mason2007}, and AM Her \citep[e.g.][]{Chanmugam1982,Chanmugam1987,Dulk1983,Mason2007}.

Recently it has been shown that the timing \citep{Koerding2008}, and sensitivity \citep{Coppejans2015}, of previous observations were insufficient to detect the radio emission. \citet{Coppejans2015} showed that CVs in a persistent high-accretion state (the novalike class) are radio emitters at a level that was below the sensitivity threshold of previous instruments. For the DN systems, none of the historical radio observations were taken in the early stages of the outburst. Through a comparison of the XRB and CV accretion states, \citet{Koerding2008} showed that radio emission produced by a transient jet should flare shortly after the rise to outburst, and then subsequently drop to undetectable levels -- explaining the lack of detections.

The XRBs show an empirical relation between the outflow properties (power, morphology) and the accretion state \citep[e.g.][]{Belloni2011,Corbel2004,Fender2004,Migliari2006,Miller-Jones2012}. As outbursts of XRBs and CVs are described by the same model and they share similar phenomenology, \citet{Koerding2008} mapped this relation onto the CVs. They predicted that DN should show a synchrotron radio flare from a transient jet on the rise to outburst, and they detected this in the DN SS Cyg. \citet{Miller-Jones2011} subsequently confirmed this behaviour in a separate outburst of SS Cyg, and \citet{Russell2016} proved that the radio outbursts of SS Cyg undergo similar evolution from outburst to outburst. Based on the timing, variability, spectral indices and brightness temperatures of the radio emission, and multi-wavelength data, \citet{Koerding2008}, \citet{Miller-Jones2011} and \citet{Russell2016} argue that the radio emission in SS Cyg is produced by a transient jet.

A number of other radio emission mechanisms have been suggested for non-magnetic CVs. Thermal emission from a large gas cloud surrounding the CV \citep[e.g.][]{Cordova1983} has been excluded in the detected CVs based on the spectral indices, brightness temperatures, or variability time-scales of the observed radio emission \citep{Fuerst1986,Koerding2008,Coppejans2015}. Non-thermal emission in the form of synchrotron, or coherent emission has also been suggested to be produced through magnetic reconnections in the disc, reflection of electrons in the magnetic field near the surface of the white dwarf, or disruption of the magnetosphere \citep{Fuerst1986,Benz1989,Benz1996}. These coherent mechanisms have been ruled out for SS Cyg \citep{Koerding2008,Miller-Jones2011,Russell2016}. The radio emission from the novalike systems is either synchrotron or coherent emission \citep{Coppejans2015}.

Only four DN (namely SU UMa, EM Cyg, TY PSc and SS Cyg) have been detected at radio wavelengths \citep{Benz1983,Benz1989,Koerding2008,Miller-Jones2011,Russell2016}. The numerous other radio observations of DN in outburst yielded non-detections with upper-limits on the order of 0.1--0.3\mJy\;\citep[][]{Fuerst1986, Benz1989, Echevarria1987}. Of the detected sources, only SS Cyg was observed with sufficient sensitivity, cadence, and timing, to look for radio emission from a transient jet.

In this paper we present 10-GHz Karl G. Jansky Very Large Array (VLA) light curves of five DN (U Gem, Z Cam, SU UMa, YZ Cnc and RX And) over the course of an outburst of each system. We detect radio emission from all five systems, proving that DN in outburst are radio emitters. Section \ref{sec:targets} gives a brief description of the targets, and our selection criteria. The observations and results are given in Section \ref{sec:observations} and \ref{sec:results} respectively.

\section{Targets}\label{sec:targets}

There are three main subclasses of DN, namely the Z Cam, SU UMa and U Gem type DN. Each class is named after its prototype.

Z Cam type DN go through intervals which show DN outbursts, and intervals (known as ``standstills'') during which the system remains at a constant optical brightness (intermediate between the quiescent and outburst levels) and does not show outbursts. According to the DIM, if the accretion rate exceeds a certain threshold, it is sufficient to maintain the accretion disc in a persistent high-state. This is the case for the novalike systems. The accretion rate of the Z Cam systems is believed to be close to this threshold, so minor accretion rate changes can shift the system in and out of standstill \citep[e.g.][]{Honeycutt1998}.

SU UMa type DN show outbursts, as well as superoutbursts. Superoutbursts are brighter, and last longer than outbursts in a given system \citep[e.g.][]{Otulakowska-Hypka2016}. Three models have been proposed to explain superoutbursts. In the \textit{Thermal Tidal Instability model} (TTI, \citealt{Osaki1989,Osaki1996,Osaki2003}), mass and angular momentum build up in the disc over successive DN outbursts and the disc radius increases. At a sufficiently large radius\footnote{The 3:1 resonance, which is approximately 0.46 times the orbital separation (see \citealt{Osaki2003} and references therein).}, the orbital period of the outer disc material resonates with the orbit of the secondary star, and the outer disc material is driven into eccentric orbits. This increases the mass transfer rate through tidal dissipation and prolongs the outburst to form a superoutburst. In the \textit{Enhanced Mass Transfer model} (EMT, \citealt{Vogt1983,Smak1984,Osaki1985}), the secondary star is irradiated during the DN outburst, which increases the mass-transfer rate and produces a superoutburst. In the \textit{Thermal-viscous Limit Cycle Instability model} \citep{vanParadijs1983,Cannizzo2010,Cannizzo2012}, superoutbursts are triggered by the same thermal-viscous instability as the DN outbursts. The instability criterion is triggered in the inner disc, and a heating wave propagates outwards as the instability is triggered in annuli at increasingly larger radii. In DN outbursts, this wave does not reach the outer disc radius. In superoutbursts, the wave extends to larger disc radii and produces a longer, brighter outburst. For more discussions on the superoutburst models see e.g. \citet{Schreiber2004}, \citet{Smak2008} and \citet{Osaki2013}. In some systems, DN outbursts are seen immediately prior to the superoutburst. These \textit{precursor outbursts} are believed to trigger the superoutburst according to the TTI and EMT models \citep[e.g.][]{Osaki2013}.

U Gem type DN show only normal outbursts. The build-up of a mass reservoir at the outer disc is prevented by the heating wave of a normal outburst reaching the outer regions every time.

\begin{table*}
  \centering
  \begin{minipage}{\textwidth}
    \caption{Properties of the target dwarf novae}
    \begin{tabular}{lllllllll}
    \hline
    Name & RA (J2000) & Dec. (J2000) & RA proper & Dec. proper & DN type & $P_{\rm orb}$ & Dist. & Incl. \\ 
    & & & motion$^{c}$ & motion & & (h) & (pc) & (deg)\\
    & & & (mas/y) & (mas/y) & &  &  & \\
    \hline
    Z Cam & 08:25:13.201 $\pm$ 0.002$^{a}$ & +73:06:39.23 $\pm$ 0.03$^{a}$ & --8.1 $\pm$ 2.5$^{a}$ & --18.0 $\pm$ 2.5$^{a}$ & Z Cam & 6.956174(5)$^{f}$ & 163 & 62\\
    RX And & 01:04:35.538 $\pm$ 0.004$^{b}$ & +41:17:57.78 $\pm$ 0.06$^{b}$ & - & - & Z Cam & 5.03743(2)$^{g}$ & 200 & 55 \\
    SU UMa & 08:12:28.264 $\pm$ 0.004$^{b}$ & +62:36:22.46 $\pm$ 0.06$^{b}$ & 7.3 $\pm$ 2.7$^{d}$ & --30.3 $\pm$ 6.9$^{d}$ & SU UMa & 1.832(1)$^{h}$ & 260 & 42\\
    YZ Cnc & 08:10:56.645 $\pm$ 0.004$^{b}$ & +28:08:33.46 $\pm$ 0.06$^{b}$ & 38 $\pm$ 8$^{e}$ & --58 $\pm$ 8$^{e}$ & SU UMa & 2.0862(2)$^{i}$ & 260 & 15\\
    U Gem & 07:55:05.235 $\pm$ 0.005$^{b}$ & +22:00:05.07 $\pm$ 0.08$^{b}$ & --26 $\pm$ 8$^{e}$ & --32 $\pm$ 8$^{e}$ & U Gem & 4.246(7)$^{j}$ & 102 & 70\\
    \hline
    \multicolumn{9}{p{17cm}}{\footnotesize{\textit{Notes:} Optical coordinates retrieved via Simbad from $^a$\citet{Hog2000}, $^b$\citet{Cutri2003}. $^{c}\mu_{\rm RA}$cos(dec). Proper motions are from $^{d}$\citet{Mickaelian2010}, $^e$\citet{Skinner2014}, $^a$\citet{Hog2000}. Orbital periods from $^{f}$\citet{Thorstensen1995}, $^{g}$\citet{Kaitchuck1989}, $^{h}$\citet{Thorstensen1986}, $^{i}$\citet{vanParadijs1994} and $^{j}$\citet{Marsh1990}. The error on the last digit is quoted in parenthesis; for example 0.2(3) is equivalent to 0.2$\pm$0.3. Distances from \citet{Patterson2011}, who estimate the uncertainty at 15-20\%. Inclinations from \citet{Patterson2011}. As a number of assumptions are necessary to determine the inclination (specifically for the low-inclination systems), the inclinations should be considered as rough estimates (see the discussion in \citealt{Patterson2011}).}}\\
    \end{tabular}
    \label{tbl:coords}
  \end{minipage}
\end{table*}

We selected targets from each of the three classes. The selection criteria for the candidate sources were that they had to be nearby ($\leq260\,$pc in \citealt{Patterson2011}) and optically bright in quiescence ($V\leq15\,$mag). This was to ensure that the American Association of Variable Star Observers (AAVSO) were able to easily monitor these targets. To guarantee that we were able to observe five systems in outburst during the observing semester, we monitored a sample of nine DN (Z Cam, RX And, SU UMa, YZ Cnc, U Gem, SY Cnc, EX Dra, EM Cyg and AB Dra). We triggered observations on Z Cam, RX And, SU UMa, YZ Cnc and U Gem. The properties of these targets are given in Table \ref{tbl:coords}, and a short description of each is included below for reference.

\subsection{Z Camelopardalis (Z Cam)}\label{sec:zcam_background}

Z Cam is the prototype for the Z Cam class of DN. It has an orbital period of 0.2898406(2)\,d \citep{Thorstensen1995}, a M$_{\rm WD}=0.99\pm$0.15 M$_\odot$ white dwarf, a mass ratio of 1.4$\pm$0.2 \citep{Shafter1983}, and a K7-type secondary star \citep{Szkody1981}. Based on the observed variability in the mid-infrared lightcurve, \citet{Harrison2014} conclude that Z Cam is a synchrotron source when it is near outburst maximum (this is discussed further in Section \ref{sec:discussion}).

All previous radio observations of Z Cam have yielded non-detections. \citet{Fuerst1986} observed it during outburst at 4.885\,GHz and obtained an upper-limit of 0.1\mJy. Further observations by \citet{Nelson1988} and \citet{Woodsworth1977} have produced 3-$\sigma$ upper-limits of 1.1\mJy\; (5\,GHz), 4.6\mJy\; (5\,GHz) and 25\mJy\; (10.6\,GHz).

\subsection{RX Andromedae (RX And)}\label{sec:rxand_background}

RX And is classified as a Z Cam type DN as it shows long standstills. In the long-term optical light curve it also shows low states that are $\sim$3.5 mag fainter than the standstills, during which time there are no outbursts \citep{Schreiber2002}. It has an orbital period of 0.209893(1)\,d \citep{Kaitchuck1989} and a K5$\pm$2 secondary star \citep{Knigge2006}.

Although RX And has been observed at radio wavelengths, there have been no prior detections. \citet{Benz1989} observed it during outburst at 1.49 and 4.86\,GHz, and obtained 3$\sigma$ upper-limits of 0.2\mJy. \citet{Nelson1988} and \citet{Woodsworth1977} also observed it and reported upper-limits of 2.6\mJy\; (at 5\,GHz), 4.6\mJy\; (at 5\,GHz) and 25\mJy\; (at 10.6\,GHz).

\subsection{SU Ursae Majoris (SU UMa)}\label{sec:suuma_background}

SU UMa is the prototype of its subclass of outbursting DN. It was discovered by \citet{Ceraski1908} and has an orbital period of 0.07635(4)\,d \citep{Thorstensen1986}. Observations at 4.885\,GHz by \citet{Fuerst1986} during the outburst maximum resulted in an upper-limit of 0.11\mJy. Effelsberg observations at 4.75\,GHz by \citet{Benz1983} resulted in a 1.3\mJy\; detection 1--2 days after the rise of the outburst, and a 0.4\mJy\; upper-limit during a subsequent quiescent epoch. Further radio observations were attempted by \citet{Nelson1988}, \citet{Echevarria1987} and \citet{Fuerst1986}, but no detections were reported. 

\subsection{YZ Cancri (YZ Cnc)}\label{sec:yzcnc_background}

YZ Cnc is an eclipsing SU UMa-type DN system with an orbital period of 0.086924(7)\,d \citep{vanParadijs1994}. Only upper-limits on the radio flux are known in the literature, despite observations during the rise and at the peak of super-outburst. Limits range from 0.11\mJy\; at 4.885\,GHz \citep{Fuerst1986}, to 0.87, 0.6 and 0.96\mJy\; at 1.5, 4.9 and 0.15\,GHz respectively \citep{Nelson1988}.

\subsection{U Geminorum (U Gem)}\label{sec:ugem_background}

U Geminorum (U Gem) is the `original' cataclysmic variable \citep{Hind1856}, and one of the best-studied systems in the class. It is a partially eclipsing system where the accretion disc is eclipsed, but the white dwarf is unobscured. It shows outbursts with a recurrence time of $\sim150$ days. Optical photometric and spectroscopic observations indicate that it develops spiral density waves in the accretion disk during outburst \citep{Groot2001_spiralshocks}. It has an orbital period of 0.1769(3)\,d \citep{Marsh1990}, and consists of a $1.20\pm0.05$ M$_{\odot}$ primary white dwarf and a $0.42\pm0.04$ M$_{\odot}$ late-type (M6) secondary star \citep{Echevarria2007}. 

The first radio observations of U Gem were attempted by \citet{Woodsworth1977} at 10.6\,GHz, which resulted in a 25\mJy\; upper-limit. Despite various attempts in the mid-1980s it was never detected at radio frequencies (neither in outburst nor quiescence). Upper-limits on the radio flux range from 1.3--2.9\mJy\; at 5\,GHz \citep{Nelson1988}, to 0.15\mJy\; at 4.885\,GHz \citep{Cordova1983}, to 0.2\mJy\; during outburst at 1.49 and 4.86\,GHz \citep{Benz1989}.

\section{Observations}\label{sec:observations}

\begin{table*}
  \centering
  \begin{minipage}{14cm}
    \caption{Observing log}
    \begin{tabular}{|l|l|l|l|l|l|l|l|}
    \hline
    Name & Obs. & VLA & Start Time & Int. & Bandpass, flux, \& & Amplitude & Polarization\\
    & & config. & (MJD) & time$^a$ (s) & polarization angle & \& phase & \& leakage\\
    & & & & & calibrator & calibrator & calibrator\\
    \hline
    Z Cam & 1 & C & 56986.11156 & 4905 & 3C48 & J0721+7120 & J0319+4130 \\
    Z Cam & 2 & C & 56987.17184 & 2007 & 3C138 & J0721+7120 & J0319+4130 \\
    Z Cam & 3 & C & 56988.14601 & 1770 & 3C48 & J0721+7120 & J0319+4130 \\
    RX And & 1 & C & 56969.40396 & 5151 & 3C48 & J0111+3906 & J2355+4950 \\
    RX And & 2 & C & 56970.35931 & 2181 & 3C48 & J0111+3906 & J2355+4950 \\
    RX And & 3 & C & 56971.37201 & 2175 & 3C48 & J0111+3906 & J2355+4950 \\
    SU UMa & 1 & C & 57027.15094 & 4902 & 3C138 & J0805+6144 & J0713+4349 \\
    SU UMa & 2 & C--CNB & 57028.06830 & 1968 & 3C138 & J0805+6144 & J0713+4349 \\
    SU UMa & 3 & C--CNB & 57029.43229 & 1995 & 3C286 & J0805+6144 & J0713+4349 \\
    YZ Cnc & 1 & C & 56984.28431 & 4635 & 3C138 & J0748+2400 & J0713+4349 \\
    YZ Cnc & 2 & C & 56985.22726 & 1854 & 3C138 & J0748+2400 & J0713+4349 \\
    YZ Cnc & 3 & C & 56987.32938 & 1860 & 3C138 & J0748+2400 & J0713+4349 \\
    U Gem & 1 & B & 57075.32906 & 4938 & 3C286 & J0805+6144 & J1407+2827 \\
    U Gem & 2 & B & 57080.15337 & 1851 & 3C138 & J0748+2400 & J0713+4349 \\
    U Gem & 3 & B & 57080.19420 & 1923 & 3C138 & J0748+2400 & J0713+4349 \\
    \hline    
    \multicolumn{8}{p{14cm}}{\footnotesize{Observations were taken at 7976--12022 MHz (X-band) with 4046 MHz of bandwidth (3-bit mode). YZ Cnc and SU UMa were in superoutburst at the time of the observations. $^a$Total integration time on source, excluding calibration scans.}}\\
    \end{tabular}
    \label{tbl:log}
  \end{minipage}
\end{table*}

DN outbursts and superoutbursts are not predictable and recur on timescales of weeks to months. Our program aimed to observe the targets immediately after the outburst rise to probe the same phase at which the radio emission flared in SS Cyg. This required the first VLA observation to be triggered during the rise from quiescence to peak outburst (which lasts approximately 24 hours). To trigger the observations at the correct time, we ran a campaign with the AAVSO. The AAVSO monitored a sample of DN in the optical and alerted us when they detected an outburst\footnote{For details of the campaign see https://www.aavso.org/aavso-alert-notice-505}. We subsequently triggered VLA observations (project VLA/14B-177) on RX And, SU UMa, U Gem, YZ Cnc and Z Cam. For triggering purposes we defined the source to be in outburst when the optical $V$-band flux rose $\sim1$\,mag above the typical flickering and orbital modulations. To ensure that we observed superoutbursts in YZ Cnc and SU UMa, we waited until superhumps or a precursor outburst were detected\footnote{The average outburst level was estimated based on the long-term optical light curves.}. 

The observing log for the observations is given in Table \ref{tbl:log}. Three separate observations were taken of each source. The first was taken to coincide as closely as possible to the end of the outburst rise. The subsequent observations were taken on the next two days. This pattern was chosen to probe the same outburst phase at which radio emission was detected from SS Cyg \citep{Koerding2008}, and to determine if our targets also showed a decline in the radio emission after the peak of outburst was reached. The deviations from this pattern were due to scheduling constraints at the VLA, or (in the case of SU UMa and YZ Cnc) to ensure that the outburst was a superoutburst.

All the observations were taken at 7976--12022 MHz (X-band) with a bandwidth of 4046 MHz (3-bit samplers). The bandwidth was split into two equal basebands, which were further subdivided into 16 spectral windows, each of which comprised 64 2-MHz channels. These observations were taken using the standard phase referencing mode; the phase and amplitude calibrator was observed for 1 minute, approximately every 4 minutes.

The data were reduced using {\sc casa}\footnote{Common Astronomy Software Applications package \citep{McMullin2007}} v4.4.0, following standard reduction techniques. Standard flux calibrators (see Table \ref{tbl:log}) and the Perley-Butler 2010 coefficients in {\sc casa} were used to set the absolute flux density scale. We did not perform any self-calibration. Two Taylor terms were used to model the frequency dependence of the sources, and Briggs weighting with a robust parameter of 1 was used. All fits were done in the image plane using the {\sc casa imfit} task and the quoted noise was determined in the vicinity of the target. Upper-limits are quoted as 3 times the noise level.

\section{Results}\label{sec:results}

\subsection{Detections}\label{sec:detections}

We detected unresolved radio emission at the positions of all five DN (four of which have not been detected previously). This increases the number of radio-detected DN by a factor of two.

\begin{figure*}
  \subfloat{
  \begin{minipage}{85mm}
    \centering
    \includegraphics[width=\textwidth]{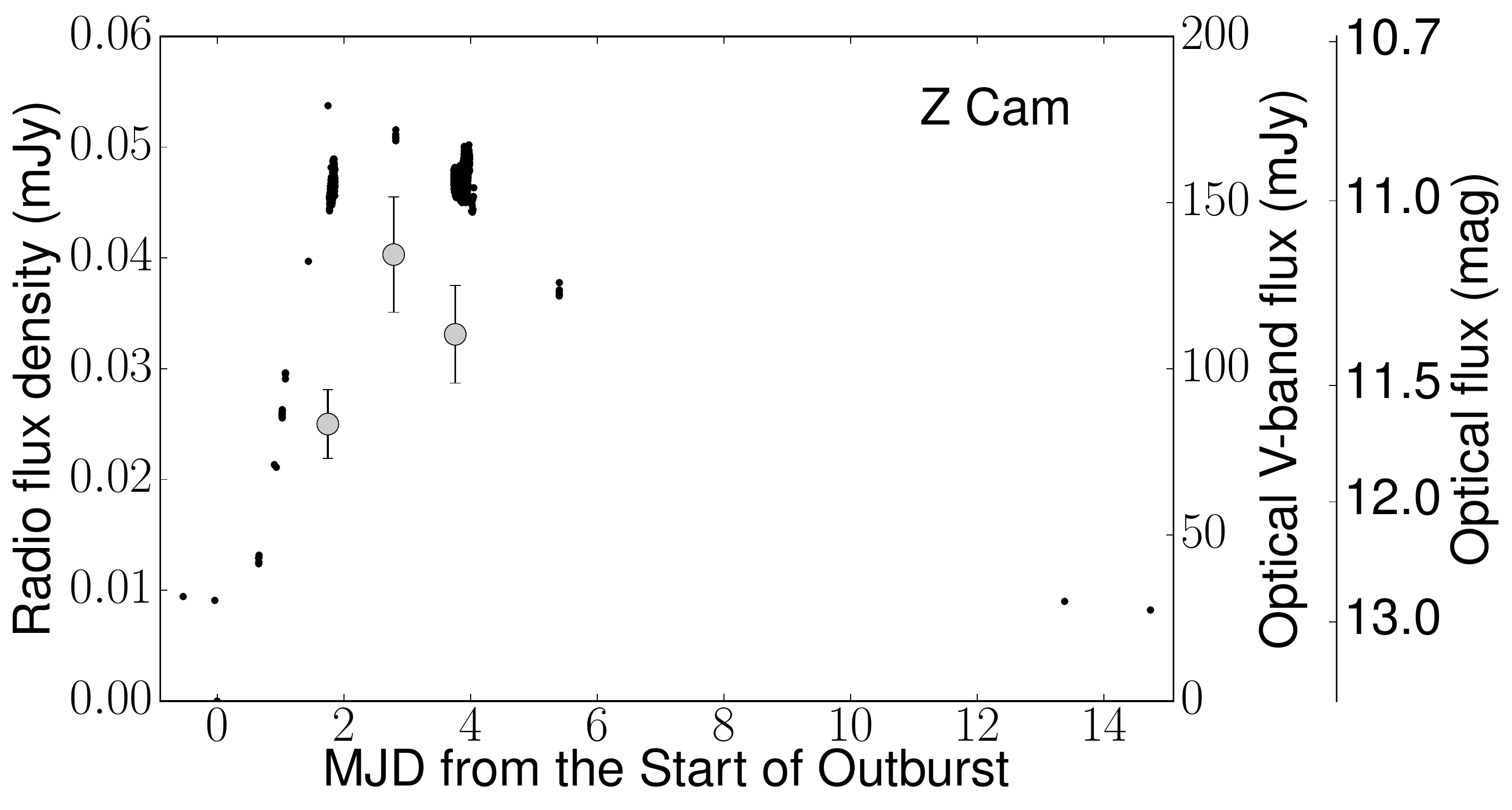}
  \end{minipage}
  \hspace{0.5cm}
  \begin{minipage}{85mm}
    \centering
    \hspace{-1.0cm}
    \includegraphics[width=0.35\textwidth]{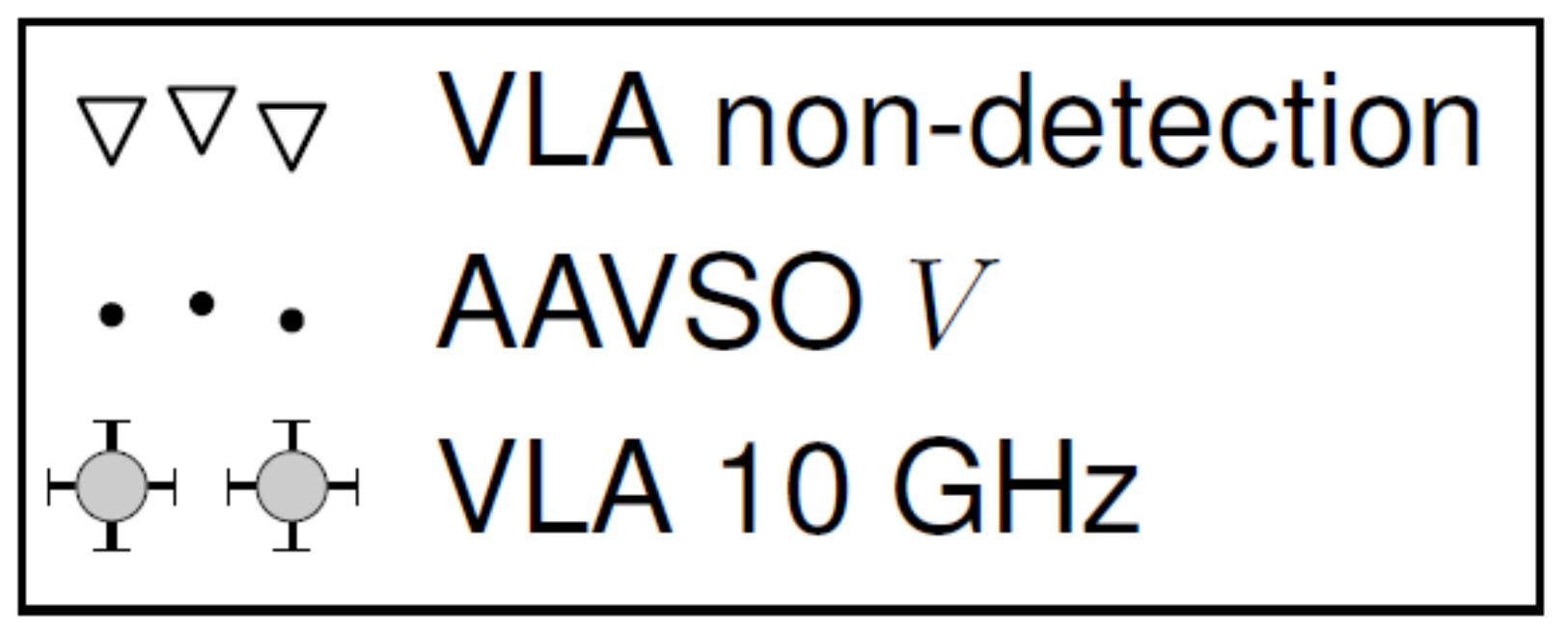}
  \end{minipage}}\\[-1ex]
  \subfloat{
  \begin{minipage}{85mm}
    \centering
    \includegraphics[width=\textwidth]{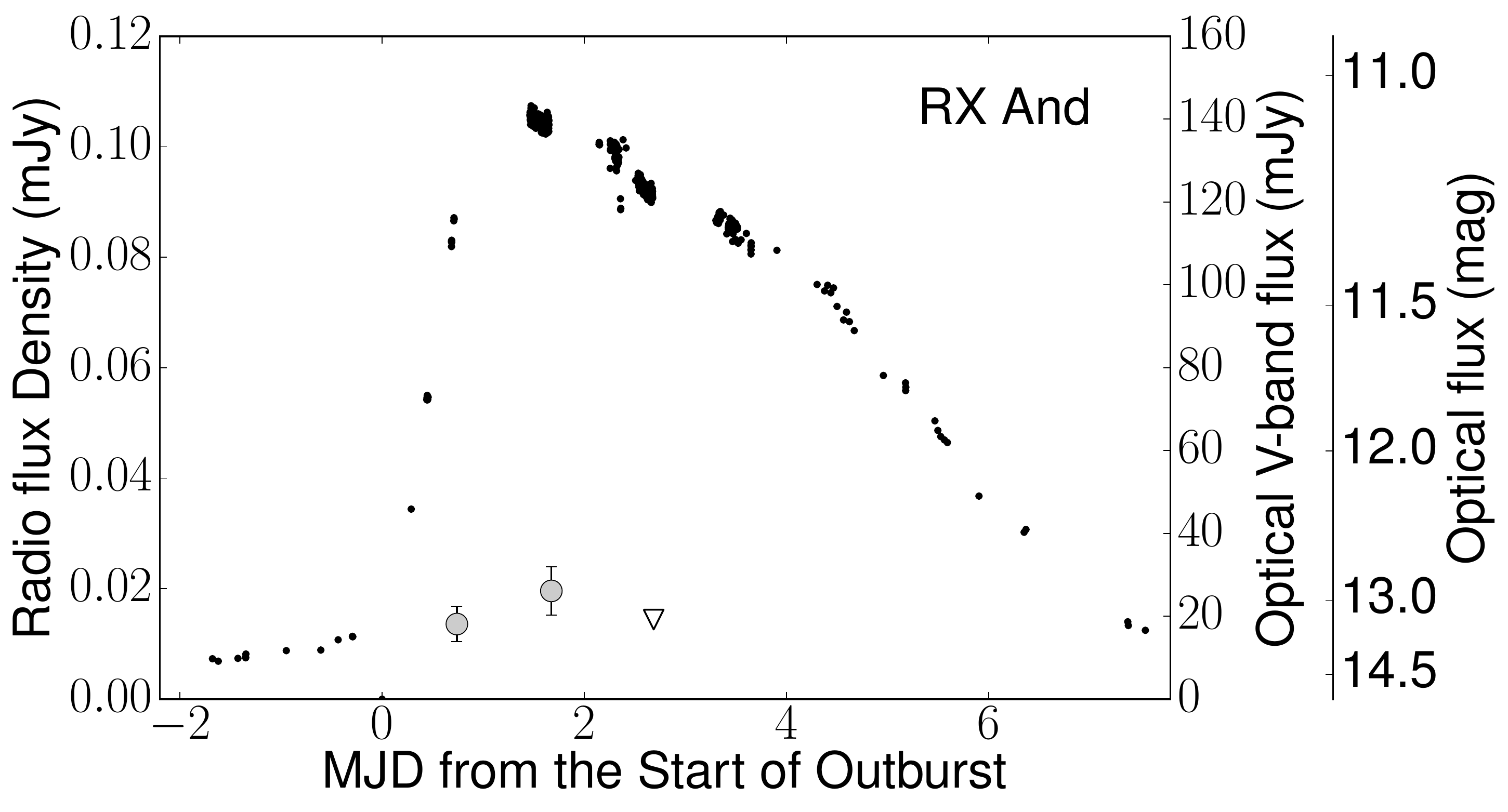}
  \end{minipage}
  \hspace{0.5cm}
  \begin{minipage}{85mm}
    \centering
    \includegraphics[width=\textwidth]{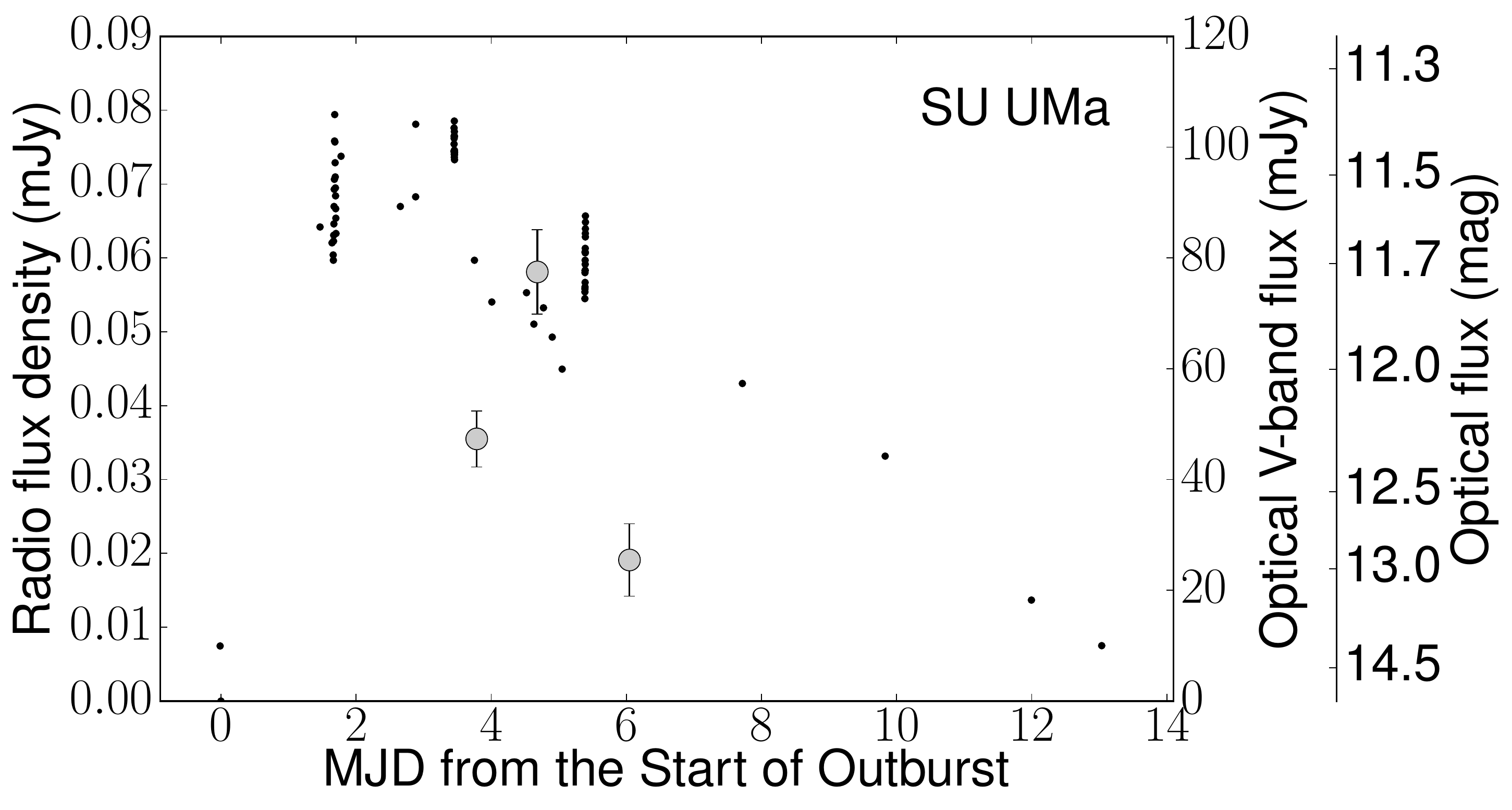}
  \end{minipage}}\\[-1ex]
  \subfloat{
  \begin{minipage}{85mm}
    \centering
    \includegraphics[width=\textwidth]{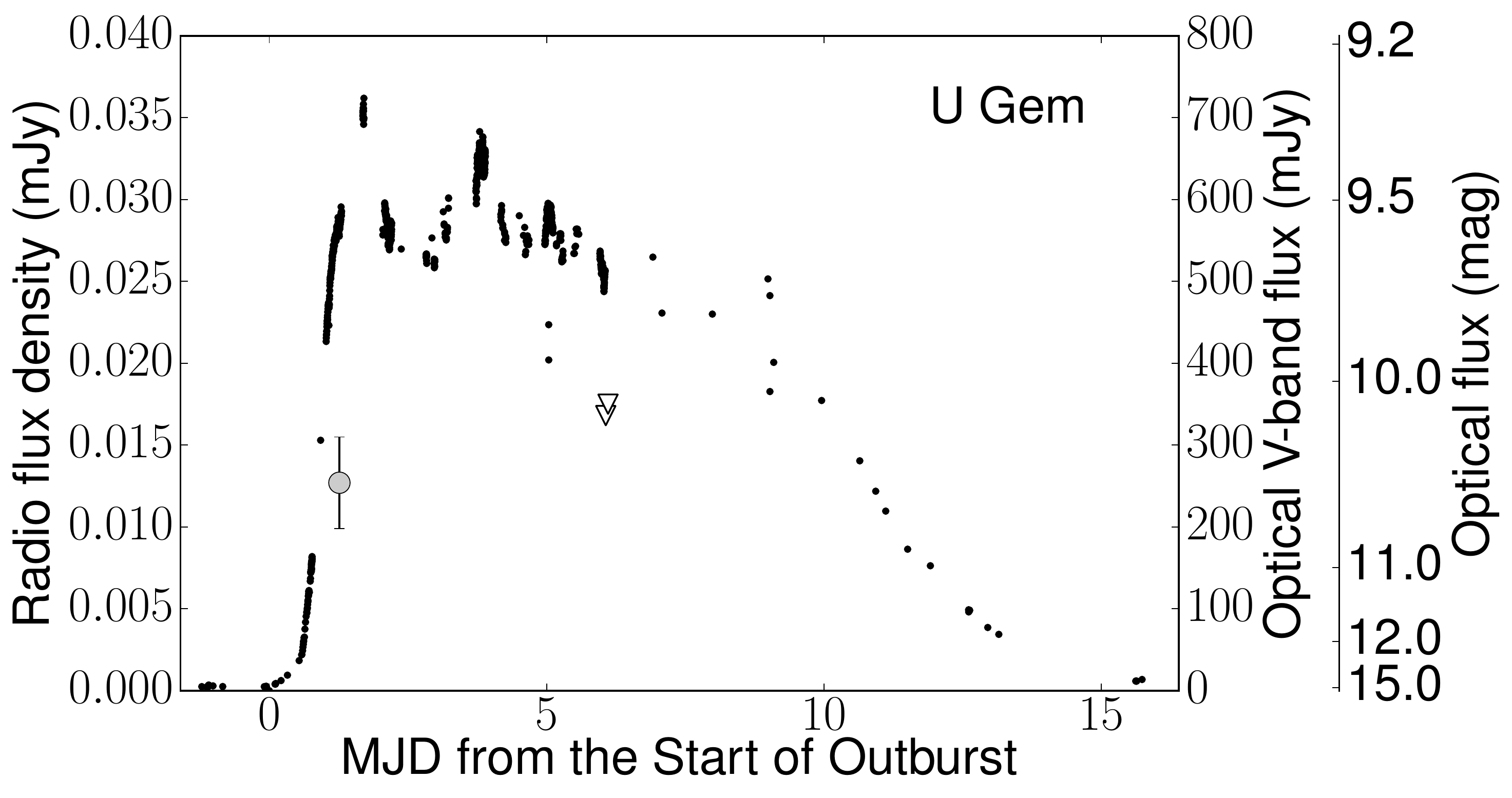}
  \end{minipage}
  \hspace{0.5cm}
  \begin{minipage}{85mm}
    \centering
    \includegraphics[width=\textwidth]{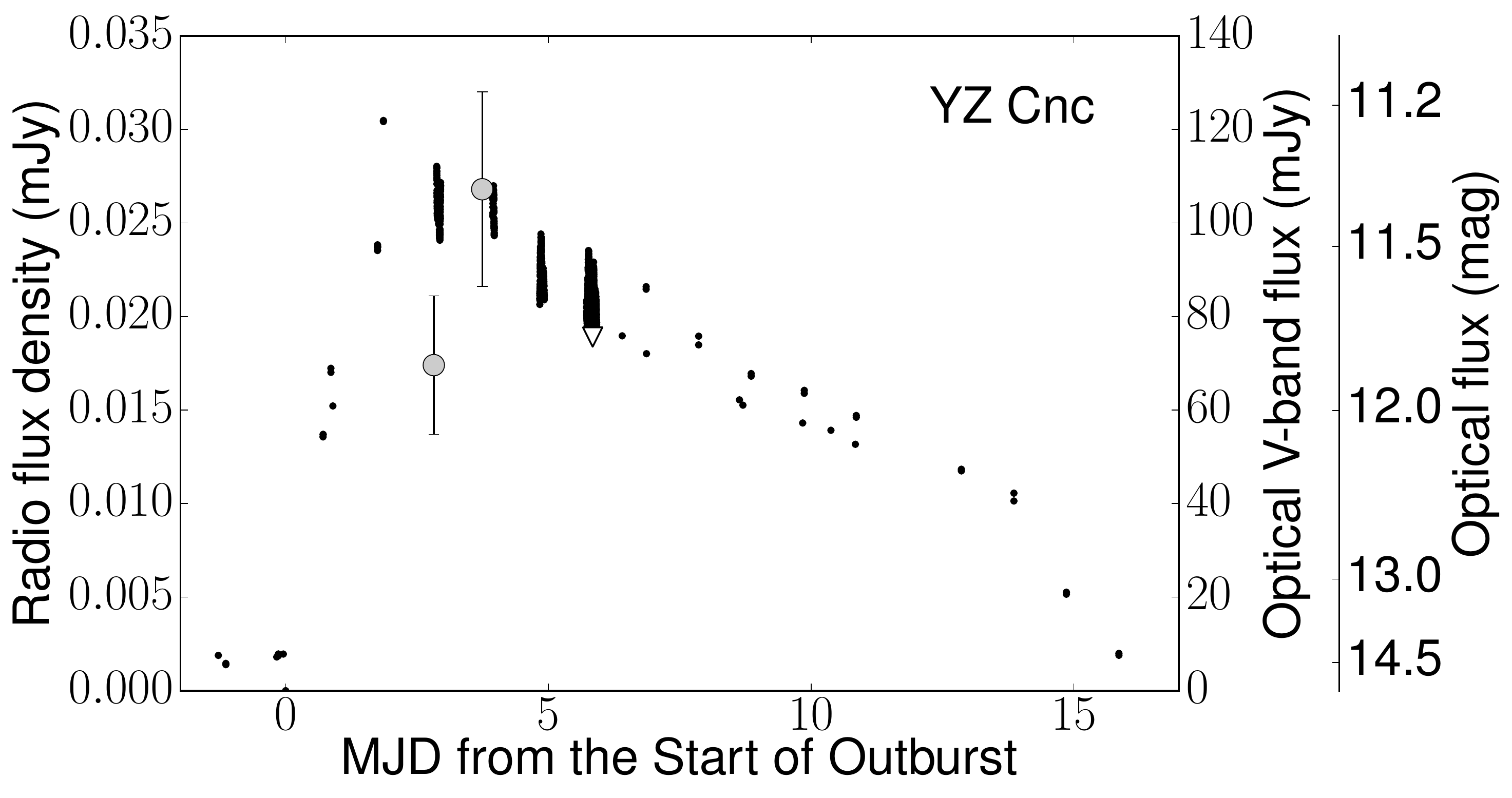}
  \end{minipage}}
  \caption{Radio (10\,GHz) and optical ($V$-band) light curves of the outbursts of RX And, U Gem and Z Cam, and super-outbursts of SU UMa and YZ Cnc. The fluxes and integration times for these light curves are given in Table \ref{tbl:results}. The precursor outburst of SU UMa has not been plotted for clarity. All optical observations are from the AAVSO International Database (see http://www.aavso.org).}
  \label{fig:lcs1}
\end{figure*}

Table \ref{tbl:results} gives a summary of the radio properties for each of the observations. Z Cam and SU UMa were detected in all three observations, and the remaining targets were detected in at least one observation. The flux densities of the detections were in the range of 20--60\muJy\; and the 3$\sigma$ upper-limits on the non-detections were on the order of 17\muJy. The radio light curves are plotted with the AAVSO optical light curves in Figure \ref{fig:lcs1}, to show the outburst phase at which they were taken.

\begin{table*}
  \centering
  \begin{minipage}{\textwidth}
    \caption{Results}
    \begin{tabular}{|l|l|l|l|r|l|l|r|c|l|}
    \hline
    Name & Outburst & Obs. & Beam Size$^b$ & PA$^c$ & RA$^d$ & Dec.$^e$ & Flux Density & CP & LP\\
    & type$^a$ &  & (arcsec$^2$) & ($\degr$) & (J2000) & (J2000) & ($\mu$Jy) & ($\mu$Jy) & ($\mu$Jy)\\
    \hline
    Z Cam & Normal & 1 & 5.21$\times$2.15 & --54 & 08:25:13.149 $\pm$ 0.029 & 73:06:39.00 $\pm$ 0.44 & 25.0 $\pm$ 3.1 & <10.2 & <6.9\\
    &  & 2 & 4.54$\times$2.08 & --71 & 08:25:13.100 $\pm$ 0.025 & 73:06:39.13 $\pm$ 0.37 & 40.3 $\pm$ 5.2 & <16.8 & <11.1\\
    &  & 3 & 4.94$\times$2.05 & --62 & 08:25:13.067 $\pm$ 0.026 & 73:06:39.34 $\pm$ 0.39 & 33.1 $\pm$ 4.4 & <16.5 & <13.2\\
    RX And & Normal & 1 & 4.49$\times$2.14 & 71 & 01:04:35.526 $\pm$ 0.030 & 41:17:57.73 $\pm$ 0.46 & 13.6 $\pm$ 3.2  & <10.5 & <6.0\\
    &  & 2 & 3.07$\times$2.19 & 87 & 01:04:35.547 $\pm$ 0.020 & 41:17:57.22 $\pm$ 0.32 & 19.6 $\pm$ 4.4 & <13.5 & <8.7\\
    &  & 3 & 3.32$\times$2.19 & 83 & - & - & <14.4 & - & -\\
    SU UMa & Super & 1 & 3.11$\times$2.16 & 75 & 08:12:28.296 $\pm$ 0.012 & 62:36:21.89 $\pm$ 0.19 & 35.5 $\pm$ 3.8 & <11.4 & <7.2\\
    &  & 2 & 5.10$\times$2.25 & --75 & 08:12:28.232 $\pm$ 0.019 & 62:36:22.07 $\pm$ 0.29 & 58.1 $\pm$ 5.7 & <17.7 & <12.0\\
    &  & 3 & 2.86$\times$1.79 & --45 & 08:12:28.310 $\pm$ 0.027 & 62:36:22.32 $\pm$ 0.40 & 19.1 $\pm$ 4.9 & <14.7 & <9.9\\
    YZ Cnc & Super & 1 & 2.93$\times$2.32 & --72 & 08:10:56.692 $\pm$ 0.021 & 28:08:32.71 $\pm$ 0.32 & 17.4 $\pm$ 3.7 & <10.5 & <6.3\\
    &  & 2 & 5.73$\times$2.25 & --65 & 08:10:56.644 $\pm$ 0.045 & 28:08:32.86 $\pm$ 0.67 & 26.8 $\pm$ 5.2 & <18.3 & <10.8\\
    &  & 3 & 2.57$\times$2.47 & --23 & - & - & <18.9 & - & -\\
    U Gem & Normal & 1 & 0.96$\times$0.65 & 64 & 07:55:05.2081 $\pm$ 0.0065 & 22:00:04.4106 $\pm$ 0.098 & 12.7 $\pm$ 2.8 & <9.0 & <7.8\\
    &  & 2 & 0.70$\times$0.60 & --3 & - & - & <16.8 & - & -\\
    &  & 3 & 0.72$\times$0.60 & --3 & - & - & <17.5 & - & -\\
    \hline    
    \multicolumn{10}{p{17cm}}{\footnotesize{Notes: All detections were consistent with point sources, and all upper-limits are $3\sigma$. $^a$`Normal' refers to a DN outburst, and `super' refers to a super-outburst. $^b$Major and minor axis of the synthesized beam. $^c$Position angle of the beam. $^e$Each radio position is consistent with the optical position and proper motion in Table \ref{tbl:coords}.}}\\
    \end{tabular}
    \label{tbl:results}
  \end{minipage}
\end{table*}

\begin{figure*}
  \subfloat{
  \begin{minipage}{85mm}
    \centering
    \includegraphics[width=\textwidth]{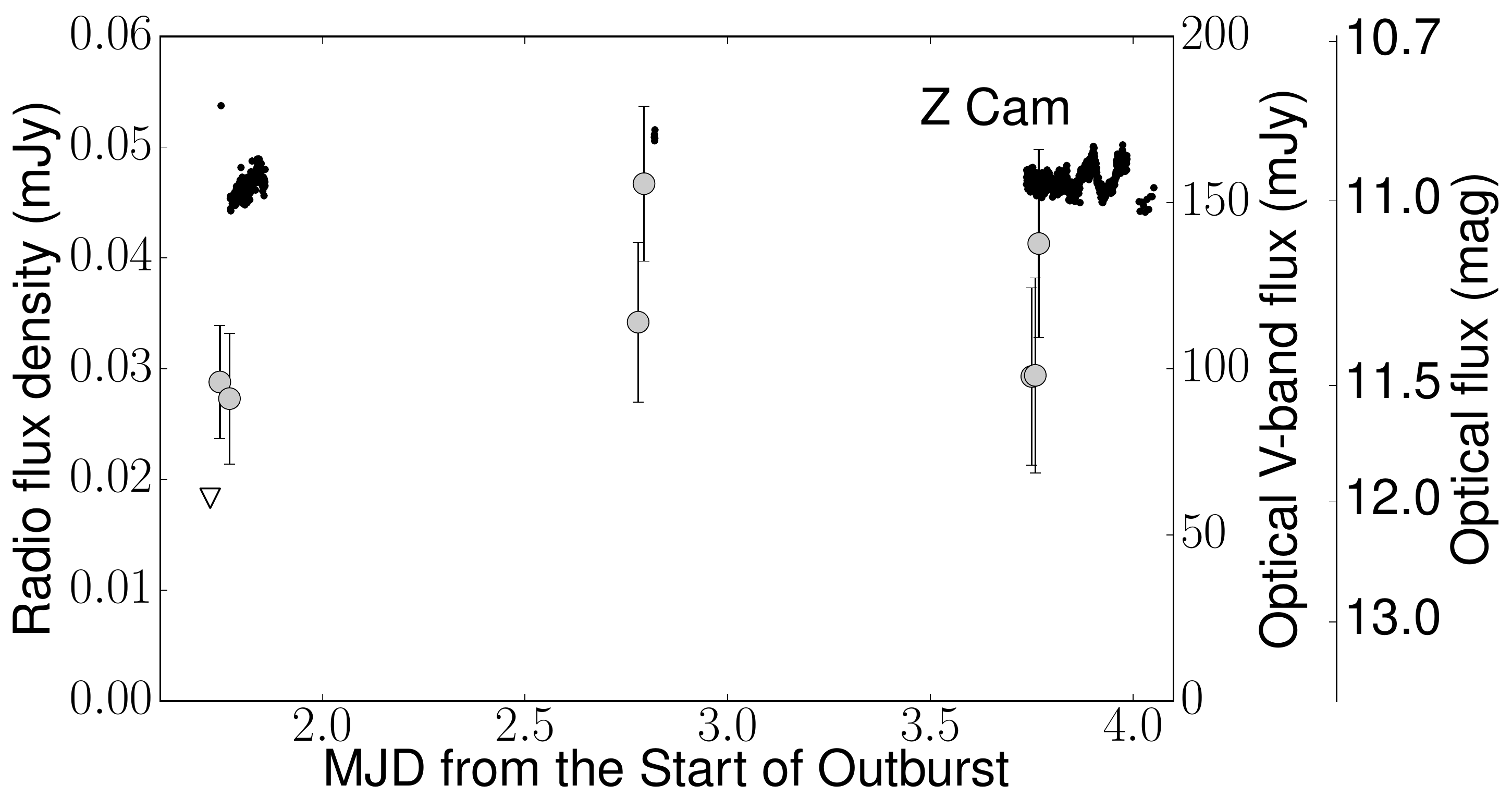}
  \end{minipage}
  \hspace{0.5cm}
  \begin{minipage}{85mm}
    \centering
    \hspace{-1.0cm}
    \includegraphics[width=0.35\textwidth]{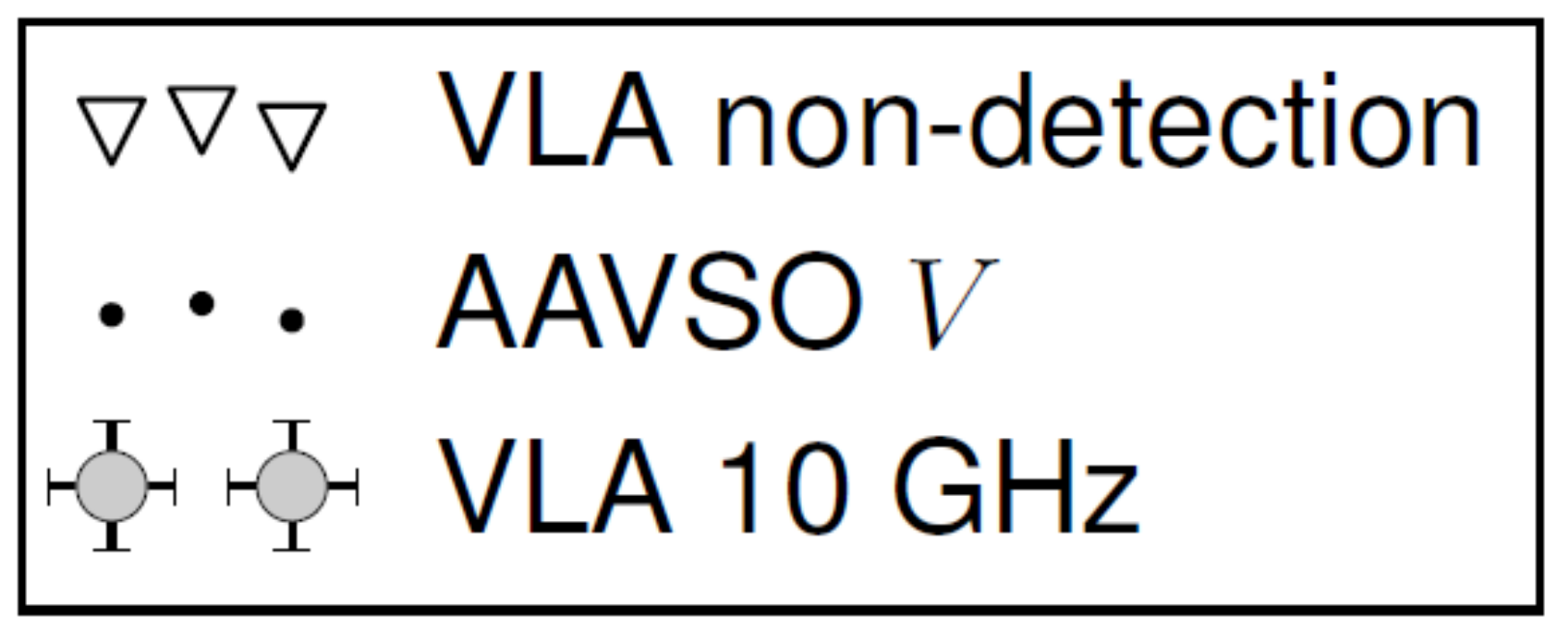}
  \end{minipage}}\\[-1ex]
  \subfloat{
  \begin{minipage}{85mm}
    \centering
    \includegraphics[width=\textwidth]{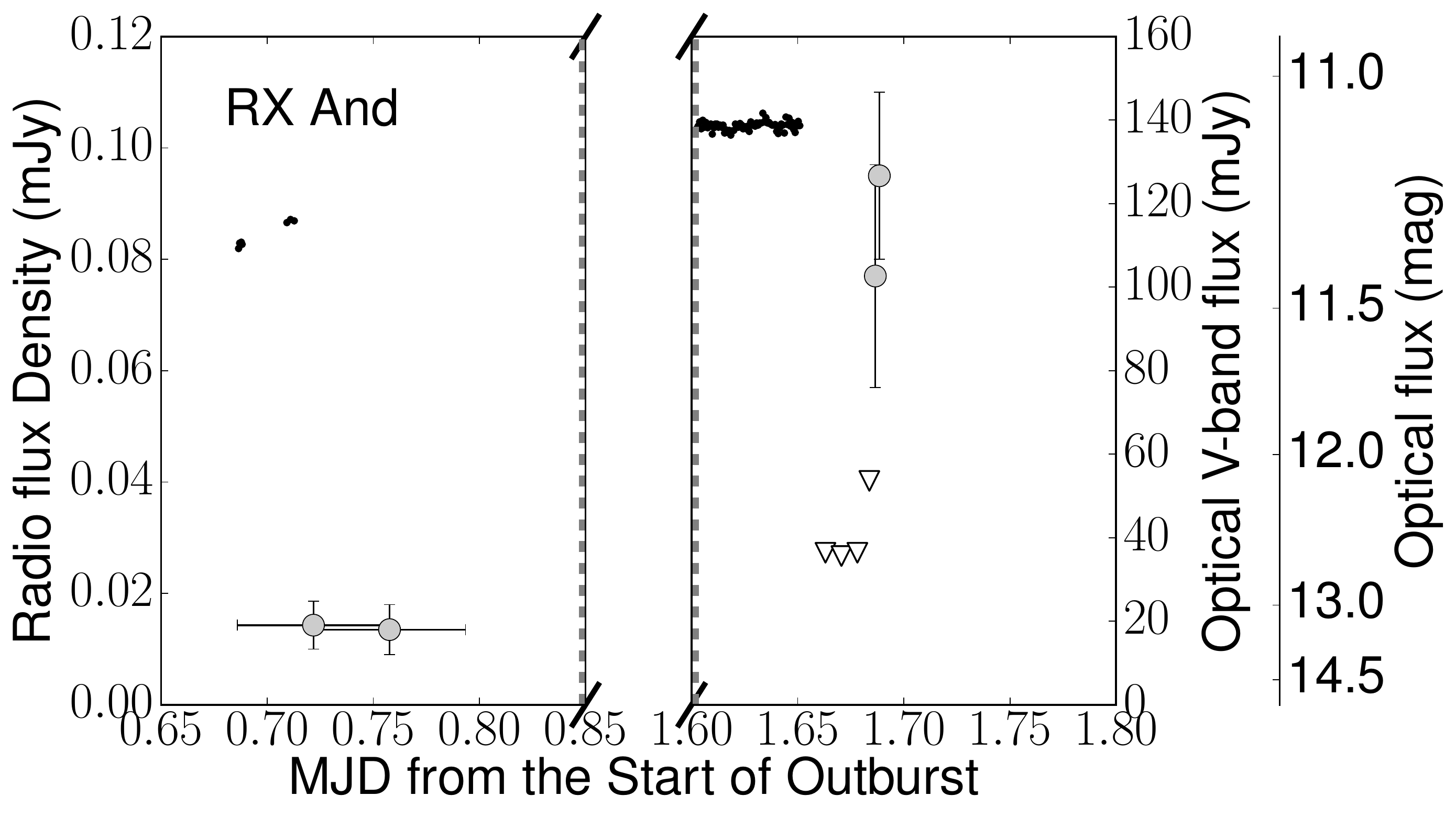}
  \end{minipage}
  \hspace{0.5cm}
  \begin{minipage}{85mm}
    \centering
    \includegraphics[width=\textwidth]{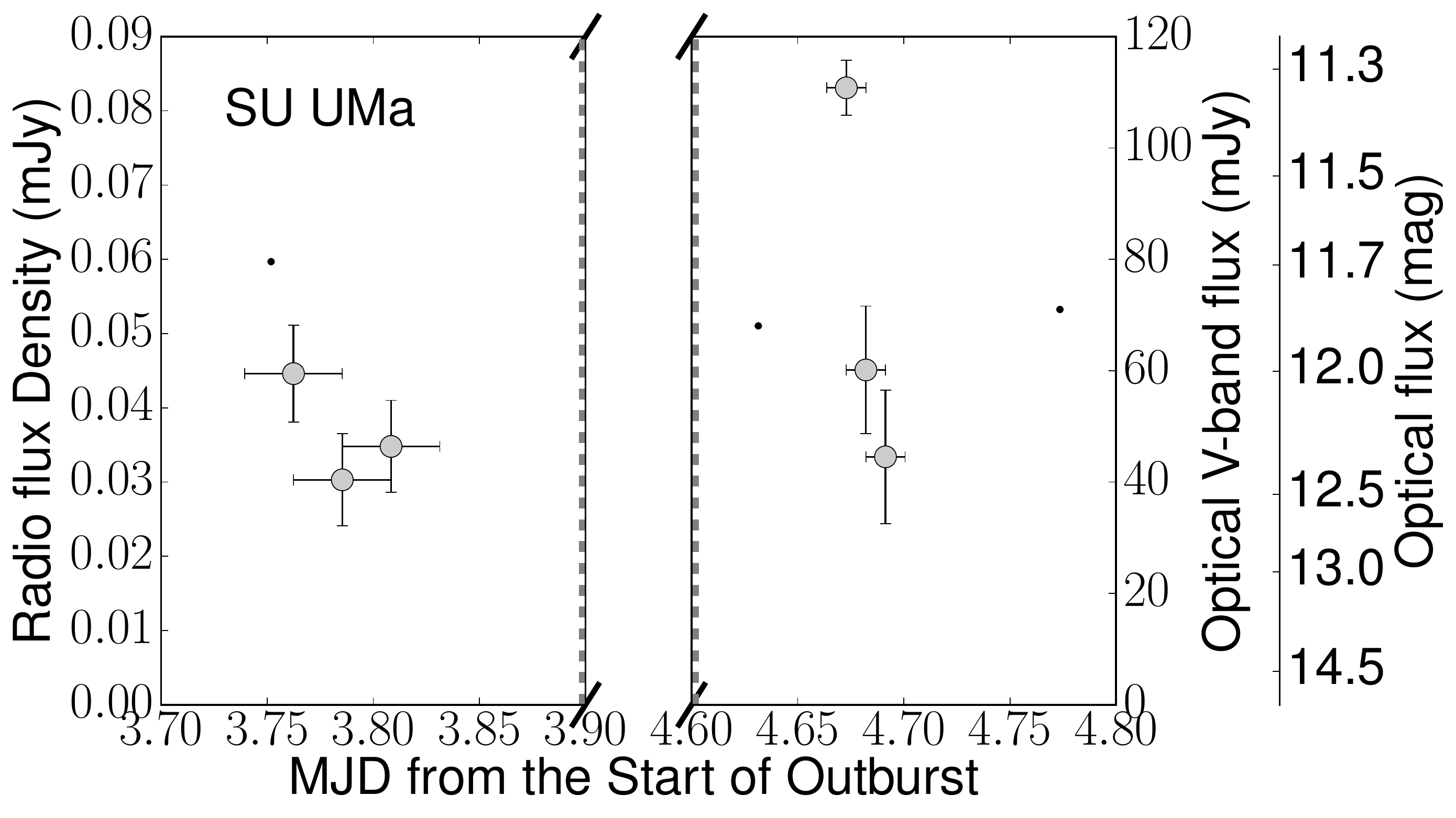}
  \end{minipage}}\\[-1ex]
  \subfloat{
  \begin{minipage}{85mm}
    \centering
    \includegraphics[width=\textwidth]{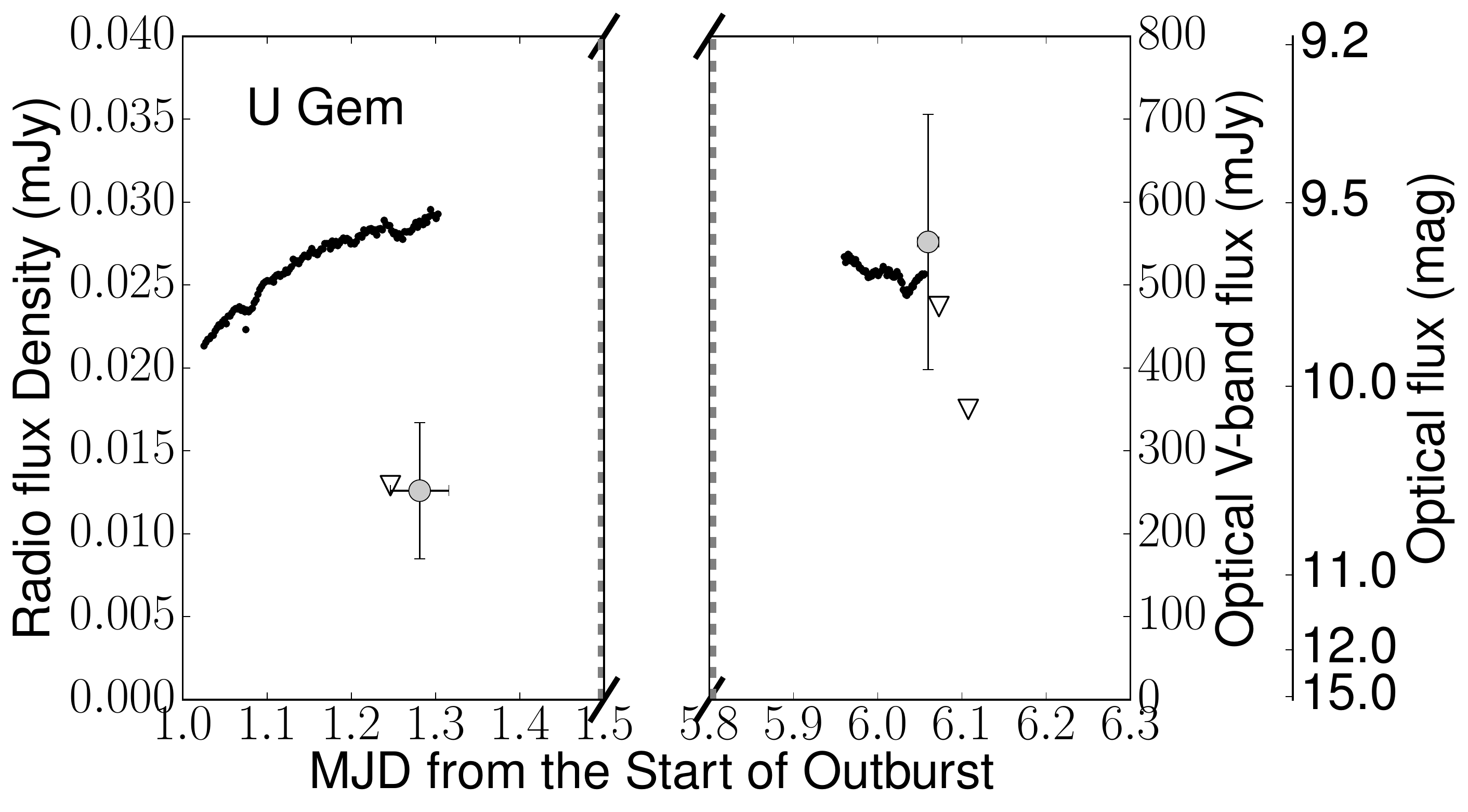}
  \end{minipage}
  \hspace{0.5cm}
  \begin{minipage}{85mm}
    \centering
    \includegraphics[width=\textwidth]{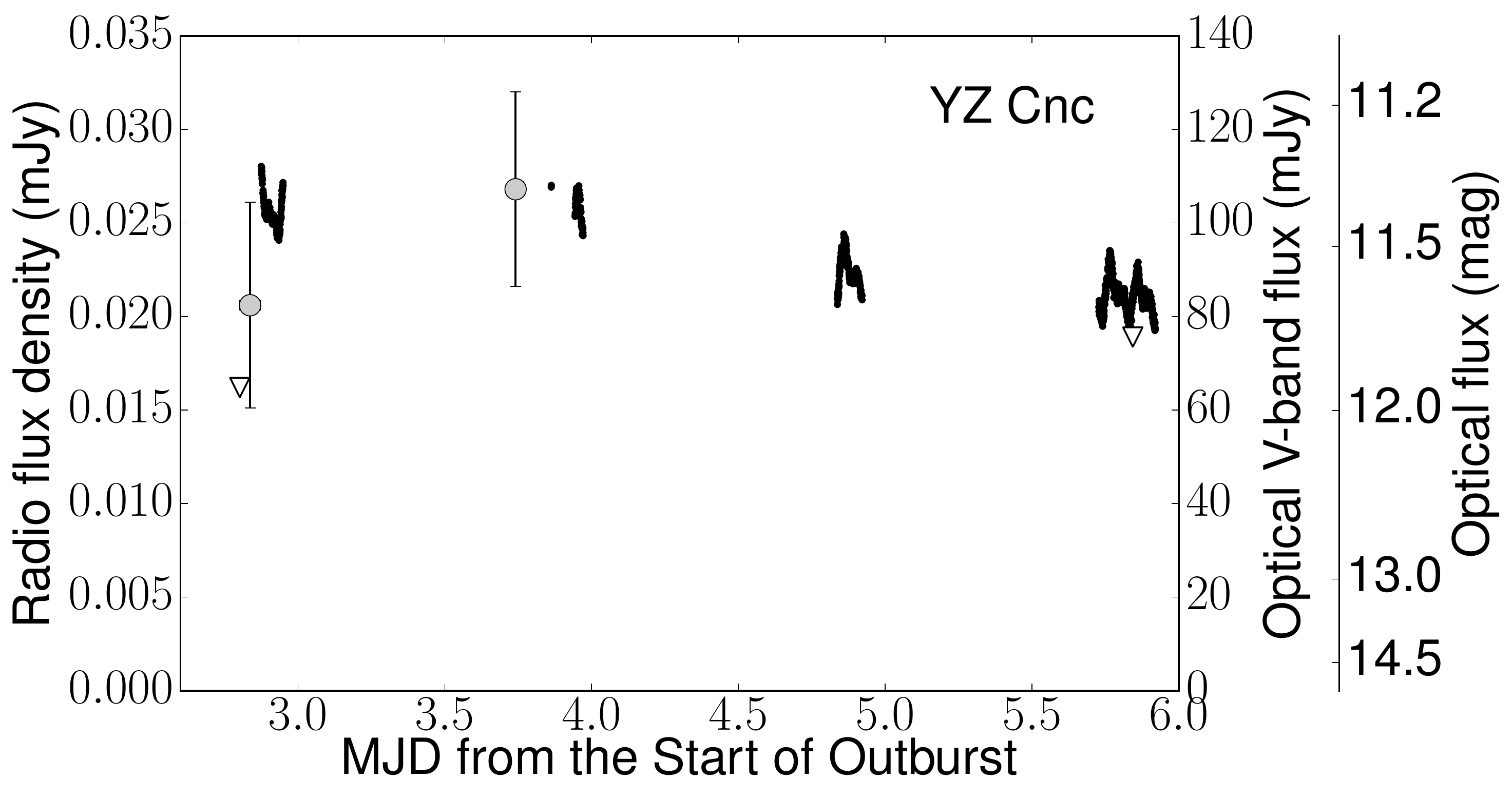}
  \end{minipage}}
  \caption{Radio (10\,GHz) observations split into shorter integrations, to show the shorter timescale variability. The fluxes and integration times for these light curves are given in Table \ref{tbl:lightcurves}. All optical $V$-band observations are from the AAVSO International Database (see http://www.aavso.org). Observation 3 of RX And and SU UMa have been omitted for clarity: The former was not detected and the latter did not show statistically significant variability.}
  \label{fig:lcs_higherres}
\end{figure*}

\subsection{Variability}\label{sec:variability}

U Gem, RX And and SU UMa all showed variability on the (approximately) one-day timescale between observations. As SS Cyg and the novalike systems showed faster-timescale variability \citep{Koerding2008,Coppejans2015,Russell2016}, we split the observations into finer segments to see if this is the case in the DN systems. The results are presented in Figure \ref{fig:lcs_higherres} and Table \ref{tbl:lightcurves}.

In RX And we detected variability on time-scales of minutes. The second observation showed a clear flare, rising from an undetectable level with a 3$\sigma$ upper-limit of 40.2\muJy, to 95.0$\pm$15.0\muJy\; in 4 minutes. As this flare occured at the end of the observation, we can not constrain its duration.

SU UMa also showed a flare, and variability on time-scales of tens of minutes, during the second observation. The flux density dropped from 83.1$\pm$8.7 to 33.4$\pm$9\muJy\; in the first 27 minutes of the observation.  

The shortest timescale on which U Gem showed significant variability was 69 minutes. It was detected at a flux density of 27.0$\pm$7.7\muJy, and in the subsequent 69 minutes dropped to undetectable levels (with a 3$\sigma$ upper-limit of 17.5\muJy).  

YZ Cnc and Z Cam were not significantly variable on timescales of minutes or days. Z Cam showed only 2-sigma variability (a 4\% chance that it was non-variable according to the variability test in \citealt{Bell2015}). According to this same test, YZ Cnc was not variable.

In all cases the signal-to-noise ratio (SNR) was too low to test for variability on timescales shorter than minutes.

\subsection{Spectral indices and Polarization}

\begin{table*}
  \centering
  \begin{minipage}{140mm}
    \caption{Spectral indices}
    \begin{tabular}{|l|l|l|l|r|c|}
    \hline
    Object & Obs. & Band & Flux Density & Spectral Index & Reduced $\chi^2$\\
    &  & (MHz) & ($\mu$Jy) & (F$\propto\nu^{\alpha}$) & \\
    \hline
    Z Cam & 1 & 7976 -- 10024 & 24.2 $\pm$ 4.0 & 0.3 $\pm$ 1.2 & - \\
    & & 9976 -- 12024 & 25.5 $\pm$ 4.7 & &\\
    Z Cam & 2 & 7976 -- 10024 & 37.2 $\pm$ 6.7 & 0.9 $\pm$ 1.2 & - \\
    & & 9976 -- 12024 & 44.6 $\pm$ 6.6 & & \\
    Z Cam & 3 & 7976 -- 9000 & 38.5 $\pm$ 9.1 & --0.1 $\pm$ 1.1 & 0.18 \\
    & & 9000 -- 10024 & 39.0 $\pm$ 9.3 & & \\
    & & 9976 -- 11000 & 32.5 $\pm$ 9.2 & & \\
    & & 11000 -- 12024 & 41.0 $\pm$ 14.0 & & \\
    RX And & 2, flare$^a$ & 7976 -- 10024 & 10.3 $\pm$ 3.3 & 4.2 $\pm$ 1.3 & - \\
    & & 9976 -- 12024 & 19.8 $\pm$ 4.2 & & \\
    SU UMa & 1 & 7976 -- 9000 & 30.7 $\pm$ 7.0 & 1.0 $\pm$ 0.9 & 2.5 \\
    & & 9000 -- 10024 & 31.7 $\pm$ 7.1 & & \\
    & & 9976 -- 11000 & 50.2 $\pm$ 6.9 & & \\
    & & 11000 -- 12024 & 32.0 $\pm$ 7.4 & & \\
    SU UMa & 2 & 7976 -- 9000 & 60.7 $\pm$ 9.2 & 0.3 $\pm$ 0.6 & 0.84\\
    & & 9000 -- 10024 & 45.9 $\pm$ 9.5 & &\\
    & & 9976 -- 11000 & 59.8 $\pm$ 9.9 & & \\
    & & 11000 -- 12024 & 65.0 $\pm$ 9.9 & & \\
    YZ Cnc & 1, 2, 3 & 7976 -- 10024 & 24.3 $\pm$ 3.6 & --2.9 $\pm$ 1.5 & -\\
    & & 9976 -- 12024 & 13.6 $\pm$ 3.7 & & \\
    \hline    
    \multicolumn{6}{p{14cm}}{\footnotesize{Spectral indices were calculated by fitting a power-law to the peak fluxes in the given frequency sub-bands. A 1$\sigma$ error is given on the spectral index. U Gem did not have sufficient SNR to calculate the spectral index, and the observations of RX And (and YZ Cnc) were combined to increase SNR. $^a$During the radio flare that occured 1.7 days into the outburst (see Figure \ref{fig:lcs_higherres}).}}\\
    \end{tabular}
    \label{tbl:spectral_index}
  \end{minipage}
\end{table*}

The spectral indices of the radio emission are shown in Table \ref{tbl:spectral_index}. Unfortunately the SNR was not high enough to constrain the spectral indices well.

None of the observations showed circular or linear polarization. The 3$\sigma$ upper-limits were on the order of 10-20\% of the measured flux (see Table \ref{tbl:results}). Consequently there are only two non-magnetic CVs from which circular polarization has been detected, namely EM Cyg and TT Ari \citep{Benz1989,Coppejans2015}. In both of these cases the circular polarization fraction was variable, and peaked at 81\% and more than 75\%, respectively.

\section{Discussion}\label{sec:discussion}

\subsection{DN in outburst as radio emitters}\label{sec:radioproperties}

We detected 10\,GHz radio emission from all five DN in outburst, with specific luminosities in the range \power{8}{13}--\power{9}{15} $\rm erg\,s^{-1}Hz^{-1}$, which correspond to luminosities of \power{8}{23}--\power{9}{25} $\rm erg\,s^{-1}$ at 10\,GHz. The radio emission is in excess of that expected from the known CV component (WD, disc and secondary star) spectra, and is variable on timescales of minutes to days in 3 of the 5 DN. The flux levels of our observations (15--80\muJy) show that the sensitivity of previous instruments was insufficient to detect these objects, as historical non-detections had 3$\sigma$ upper-limits in the range 0.1--0.3 mJy. All three classes of DN (U Gem, Z Cam and SU UMa-type DN) were represented in this sample, as well as systems in outburst and superoutburst. These results show that DN in outburst are radio emitters.

The maximum flux that SU UMa reached in our observations was 83.1$\pm$8.7\muJy, which is significantly fainter than the 1.3\mJy\footnote{Based on Figure 2 in \citet{Benz1983} we judge the noise to be $\sim$230\muJy/beam} detection of SU UMa in outburst by \citet{Benz1983}. \citeauthor{Benz1983} observed SU UMa with Effelsberg at 4.75\,GHz in two separate outbursts at 1--2 days after the optical peak (during the decline), and in quiescence. The combined outburst observations gave a detection at 1.3\mJy, and the quiescent observation produced an upper-limit of 0.4\mJy. Given the large discrepancy between the outburst flux measurements, we looked for sources of confusion within the 2.4$\arcmin$ Effelsberg beamwidth in our higher resolution VLA images. One source\footnote{at RA$_{\rm J2000}=$08:12:26, Dec$_{\rm J2000}=$+62:36:36} (an unclassified object) with a flux density of 33$\pm$9\muJy\;was within this beam, at 0.34$\arcmin$ away from SU UMa. To produce a 1.3\mJy\; detection, this object would either need to have decreased in flux by a factor $\sim$40 since 1982\footnote{Assuming a flat spectral index between 4.75 and 10\,GHz, and a gaussian fall-off in beam sensitivity}, or have a spectral index of --3.7$\pm$0.2 ($\rm{F_{\nu}\propto\nu^{\alpha}}$). This source does not show variability on time-scales of days in our observations, but this does not eliminate the possibility that it was variable at the time of the Effelsberg observations. Nearby sources outside the beam were not sufficiently bright to account for the discrepancy. SU UMa was variable in our observations, and high-amplitude flares have been detected in SS Cyg \citep{Koerding2008,Miller-Jones2011,Russell2016}. It is therefore possible that SU UMa could have flared to 1.3\mJy, but as the Effelsberg detection was made in observations averaged from two separate outbursts, a flare should have been averaged out in these data. A decrease in flux over the decades between the observations could also explain the discrepancy, but the non-detections by \citet{Fuerst1986}, \citet{Nelson1988} and \citet{Echevarria1987} argue against this. It is also possible that our VLA observations resolved out some of the flux, as Effelsberg samples different spatial scales. Based on these arguments we cannot confirm, or refute, the validity of the historical 1.3\mJy\; outburst detection of SU UMa.

It is not yet clear whether DN are radio emitters in quiescence. The only CV for which we have high-sensitivity radio observations during quiescence is SS Cyg, and it was not detected down to a 3-sigma upper-limit of 89\muJy\; \citep{Koerding2008}. All three historical radio detections (SU UMa, EM Cyg and TY PSc) were made during outburst, so there is circumstantial evidence indicating that the radio emission during outburst is brighter. High-sensitivity observations of quiescent DN are necessary to determine if the radio luminosities are fainter in quiescence.

\subsection{Radio emission mechanism in DN outbursts}\label{sec:emission_mech}

As discussed in Section \ref{sec:intro}, various mechanisms to produce synchrotron, thermal and coherent radio emission in non-magnetic CVs have been suggested. In the novalikes, the radio emission is non-thermal, and is consistent with synchrotron or coherent emission \citep{Coppejans2015}. In the DN SS Cyg, the radio emission in outburst is explained as synchrotron emission from a transient jet \citep{Koerding2008,Miller-Jones2011,Russell2016}. We now discuss what type of radio emission is seen in our sample of DN.

The emission region would need to have a radius that is $\sim 10^{2}-10^{3}$ larger than the orbital separation to produce the flux densities we observe, if it is optically thick thermal emission from an ionised gas at a typical brightness temperature of 10$^4$--10$^5$\,K. In this sample the smallest possible emission region would therefore be on the order of \power{1}{13}\,cm. Wind speeds in CVs have been measured up to 5000\,km\,s$^{-1}$ \citep[e.g.][]{Kafka2009}. Even assuming an extremely fast wind of speed $10^{4}$\,km\,s$^{-1}$, the shortest time on which changes can be propagated over the emission region, is approximately 170 minutes. As we observe variability on timescales down to 4 minutes in this sample, the observed radio emission can not be optically thick thermal emission.      

If it is optically thin thermal emission, it is unlikely to be reprocessed optical radiation from the CV (e.g. by a surrounding gas cloud), as the radio light curves do not follow the optical light curves. Following the arguments in \citet{Koerding2008}, the radio emission would need to be produced directly by an outflow (either a wind or jet). The upper-limit on the mass accretion rate in DN in outburst is $\sim10^{-8}M_{\odot}$\,${\rm y^{-1}}$, so this sets the maximum mass flow rate of the outflow (if none of the matter were accreted\footnote{and no additional matter from the WD were to be ejected}). From Equation 8 in \citet{Wright1975}, the upper-limit on the optically thin thermal flux density from a uniform velocity wind (with a speed on the order of $10^3\,\rm{km\,s^{-1}}$) with this flow rate is $\sim$1\muJy\;. Even if all the accreted material were carried off in a wind, the optically thin thermal emission would be insufficient to produce the flux densities we observe.

Synchrotron emission or coherent emission are both consistent with the observations, but as we detected no circular or linear polarization down to 3-sigma upper-limits of $\approx$10\%, we conclude that the observed radio emission is more likely to be synchrotron. The spectral indices are not sufficiently well-constrained in these observations to confirm this, but are consistent with optically thick or thin synchrotron emission.

For Z Cam there is additional evidence indicating that this is synchrotron emission. \citet{Harrison2014} found that Z Cam is a synchrotron source at mid-infrared frequencies, based on observations from the WISE mission \citep[Wide-field Infrared Survey Explorer,][]{Wright2010}. Near the peak of the visual outburst, they detected rapid variability with an amplitude of $\sim2.5$\mJy\;at 12$\,\mu$m ($W3$ band), and only marginal variability at shorter wavelengths. Based on this, they concluded that the emission is most likely a synchrotron jet.

Jets are observed in the form of self-absorbed compact jets, or as discrete ejection events: The former have flat spectral indices, and the latter progress from inverted to steep spectral indices as the ejecta becomes optically thin. Depending on the type of ejection, and the timing of the observations, synchrotron emission from a jet can produce a range of spectral indices (like that of our sample). The flux density of the 12$\,\mu$m emission in Z Cam \citep{Harrison2014} peaked at 5.0$\pm$0.5\,mJy. In comparison, our 10\,GHz radio observations were significantly fainter, peaking at a flux density of 0.0467$\pm$0.0007\,mJy. If the 12 $\mu$m variability of $\sim2.5$\mJy\;corresponds to a rise in the jet emission, this would suggest that the spectral index must be relatively inverted, $\alpha\gtrsim 0.6$ (note that the observations were not simultaneous). Spectral indices of 0.6--0.8 have been seen at some epochs for the synchrotron jets of MAXI J1836$-$194 \citep{Russell2013}. \citet{Harrison2014} also report a potential $160\pm50$\mJy\;detection with IRAS at 12\,$\mu$m at the peak of a Z Cam outburst. If the marginal IRAS detection is real, the spectral index could be even more inverted. Given both the variable nature of the source, and the potential for a spectral break between radio and mid-IR wavelengths, we caution against over interpreting the spectral index.

\begin{figure*}
  \centering
    \includegraphics[width=\textwidth]{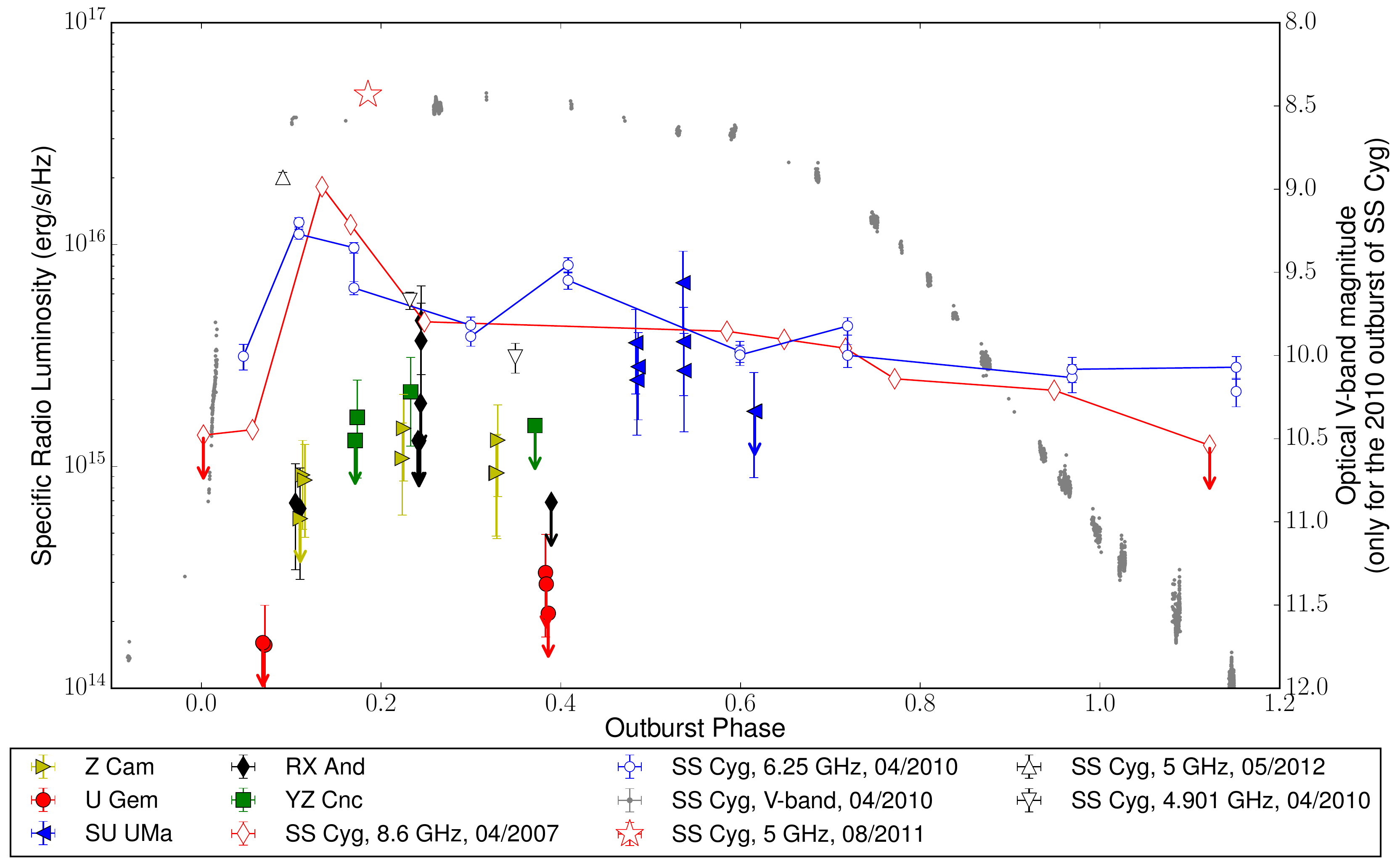}
    \caption{Specific radio luminosity of DN as a function of outburst phase. For comparison purposes, the start of the outburst (phase 0) is defined as the point on the rise to outburst at which it is $V=1\,$mag brighter than in quiescence, and the end (phase 1) is the equivalent point on the decline. In the two superoutbursts (SU UMa and YZ Cnc), phase zero is defined during the precursor outburst. The radio observations of SS Cyg are from \citet{Koerding2008,Miller-Jones2011,Miller-Jones2013} and \citet{Russell2016}. For clarity, not all previous radio observations of SS Cyg are plotted; the full light curve is in \citet{Russell2016}. Note that \citeauthor{Russell2016} set phase 0 of the outburst to the time at which SS Cyg reached $V=10$\,mag (which is 2\,mag brighter than the quiescence level), so our definitions of the outburst phase differ. We do not use the same phasing definition here, as the outburst magnitudes and amplitudes differed between the sources. The optical $V$-band data for SS Cyg are from the AAVSO international database. Note that the optical axis only refers to the V-band data for the 2010 outburst of SS Cyg; it does not give the simultaneous optical magnitude for any other outburst. For a colour version of this figure, please see the online material.}
  \label{fig:outburstphase}
\end{figure*}

\begin{figure}
  \centering
    \includegraphics[width=0.48\textwidth]{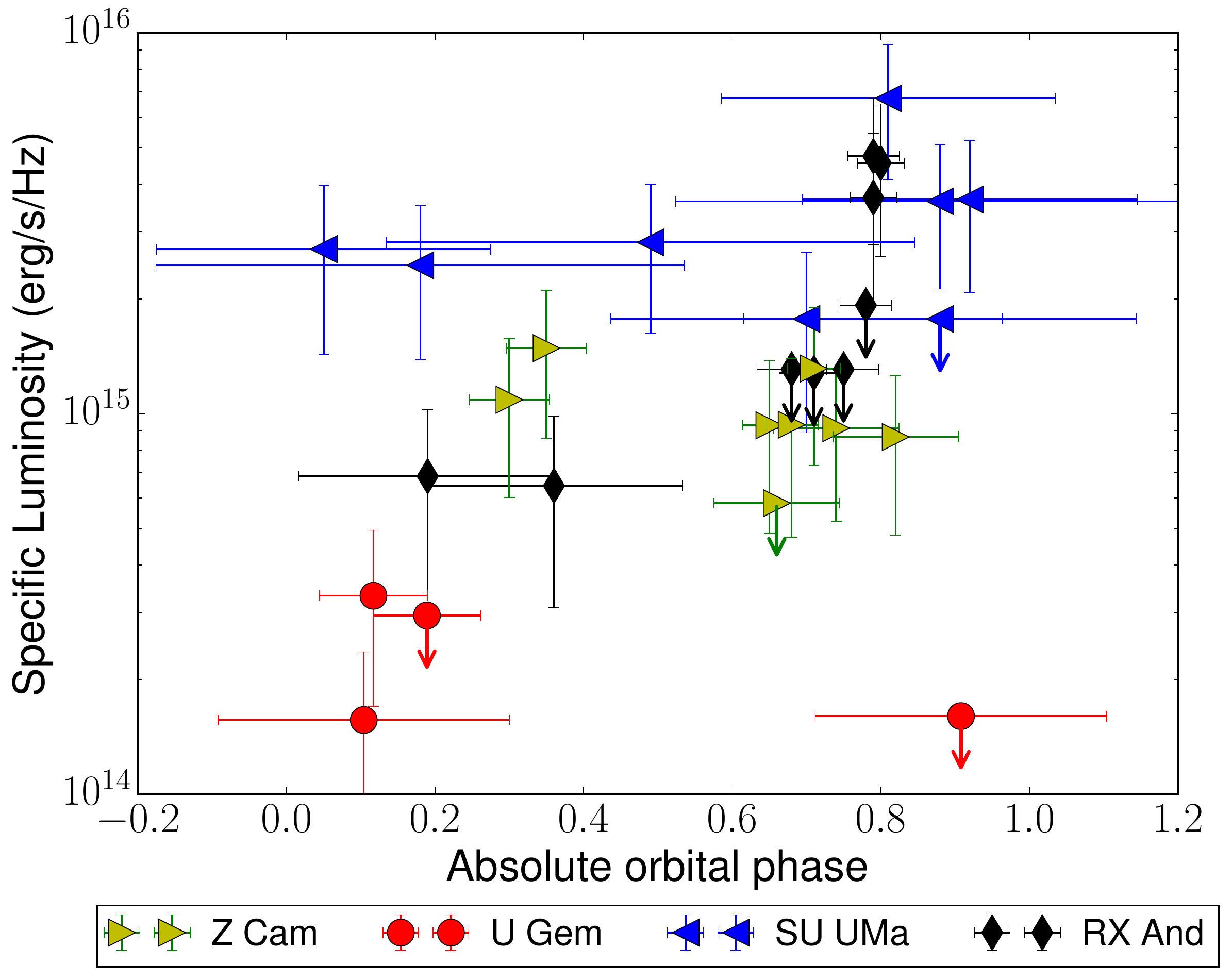}
    \caption{The radio luminosity phased according to the orbital period, using the ephemerides referenced in Table \ref{tbl:lightcurves}. As the radio luminosity does not show a dependence on orbital phase, the variability cannot be explained by an eclipse of the emission region. An orbital phase of zero corresponds to inferior conjunction, and the indicated phase error takes the integration time of the observation into account. Note, that as only the relative phase is known for YZ Cnc, it is not plotted here. For a colour version of this figure, please see the online material.}
  \label{fig:phase}
\end{figure}

For the DN SS Cyg (a U Gem-type DN), the synchrotron emission is produced by a transient jet \citep{Koerding2008, Miller-Jones2011, Russell2016}. \citet{Russell2016} compared the radio outbursts of SS Cyg over different outbursts and showed that the radio behaviour does not vary significantly between outbursts. The radio luminosity peaked on the rise to outburst (within 0.5-2 days after SS Cyg reached $V=10$ mag), decreased over the course of the outburst and was subsequently not detected in quiescence. Near the peak of an outburst, extreme radio flares (in excess of the already increased radio emission during the outburst) were detected. Based on the analogy between the outburst states in DN and XRBs, this behaviour is characteristic of a jet \citep{Koerding2008}.

Figure \ref{fig:outburstphase} shows the radio lightcurves of our sample of DN with those of SS Cyg, as a function of the outburst phase. We now compare and contrast the radio lightcurves of the DN to SS Cyg.

In three of the DN we observed variability on time-scales of days to minutes. This is consistent with SS Cyg, which has been observed to vary on timescales of days down to 30 minutes \citep{Russell2016}. The orbital period of DN is on the order of hours, and orbital modulations on shorter timescales are observed at other wavelengths in CVs. If the radio emission in the DN were to show an orbital phase dependence, it would isolate the emission region in the binary. Figure \ref{fig:phase} shows the specific radio luminosity phased according to the orbital period ($P_{\rm orb}$). As was the case for SS Cyg, the emission shows no orbital phase dependence.

SS Cyg shows increasing radio emission on the rise to optical outburst, followed by a radio flare that lasts a few hours \citep{Russell2016}. We did not detect a high-amplitude radio flare like that of SS Cyg in our observations, but we could have missed it given our sampling cadence, as our observations were 1--2 hours long and separated by a day or two (see Table \ref{tbl:log}). U Gem was observed at a corresponding earlier phase in the outburst, and was then observed 5 days later. If the radio emission in U Gem followed the same template as SS Cyg, then the sampling could also explain the low luminosity. Note that we do not consider the brightening in RX And at 1.7 days into the outburst (phase 0.25) to be equivalent to the high-amplitude flare in SS Cyg. It has a comparatively lower amplitude to SS Cyg (although this could be explained by sampling), but more importantly it occurs later in the outburst. From simultaneous X-ray observations, \citet{Russell2016} found that the radio flare in SS Cyg was coincident with the initial disc material hitting the boundary layer. This condition occurs before phase 0.25 in the outburst.

For a full comparison to SS Cyg, higher-cadence radio observations throughout the outburst are necessary. From these observations it is clear that SS Cyg is not unique in the type of radio emission, the luminosity (in the plateau phase\footnote{After phase 0.3 in Figure \ref{fig:outburstphase}}), or the variability time-scales. Higher cadence observations during the rise phase, and during later phases, of the outburst are still needed to test the extent to which the radio emission in DN follows the SS Cyg synchrotron-jet template.

\subsection{What determines the radio luminosity?}\label{sec:radio_scale}

\begin{figure}
  \centering
    \includegraphics[width=0.48\textwidth]{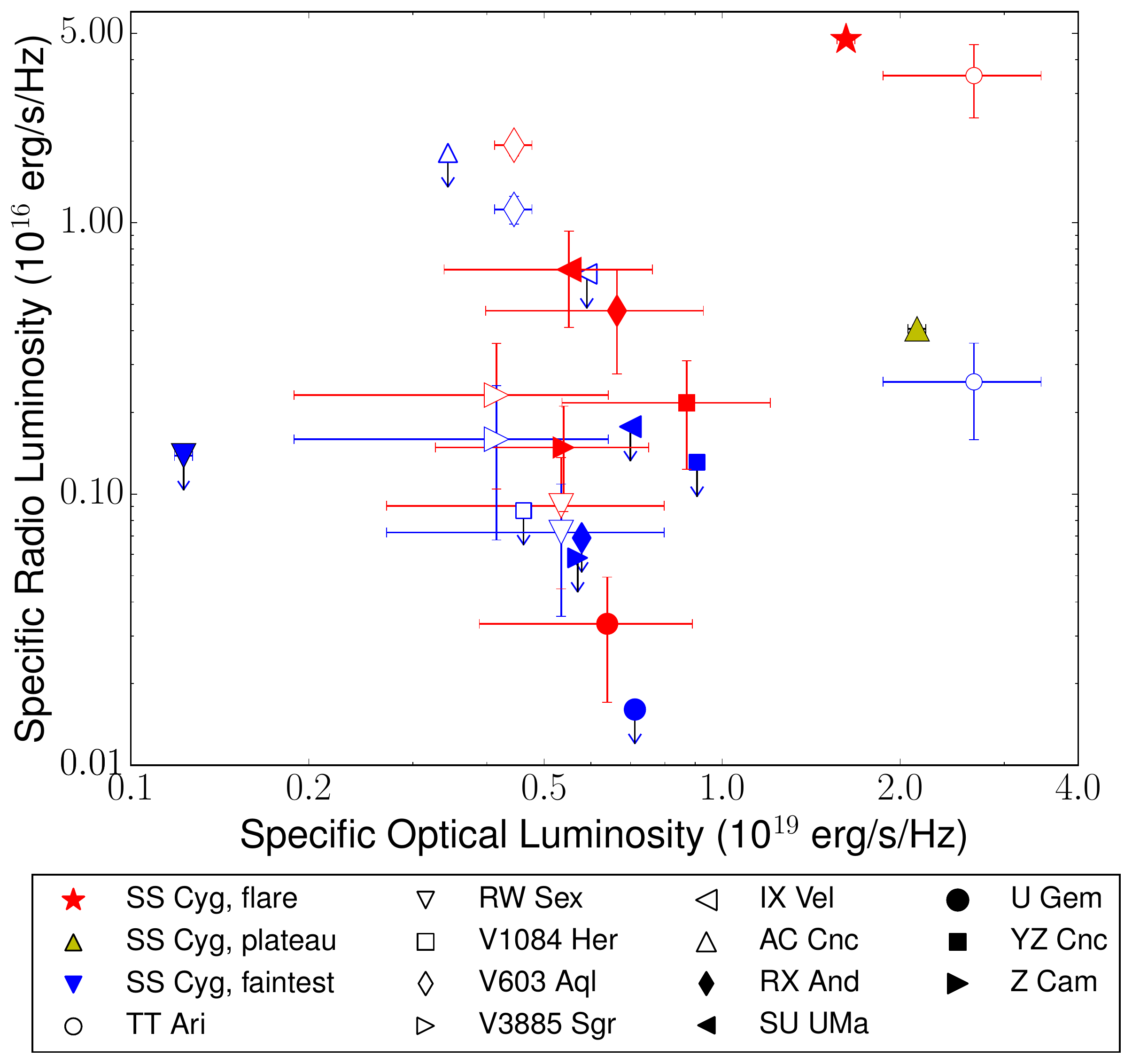}
    \caption{Distribution of the radio and quasi-simultaneous optical luminosities of all recent, high-sensitivity observations of non-magnetic CVs. The DN (solid symbols) and novalikes (empty symbols) occupy the same phase space. As most of the sources are variable, we plot only the brightest (red) and faintest (blue) radio detection (or most constraining upper-limit) for each source. The optical luminosities (from the AAVSO international database) are quasi-simultaneous to the radio observations. The brightest detection of TT Ari was during a flare with a circular polarization fraction of more than 75\%, which is in contrast to the other sources which had circular polarization fraction upper-limits of $\sim$10\%. Radio fluxes were taken from \citet{Miller-Jones2011}, \citet{Koerding2008} and \citet{Russell2016} (SS Cyg), \citet{Coppejans2015} (V1084 Her, RW Sex, V603 Aql and TT Ari) and \citet{Koerding2011} (V3885 Sgr and IX Vel). A colour version of this figure is available online.}
  \label{fig:radiolum_opticallum}
\end{figure}

\begin{figure}
  \centering
    \includegraphics[width=0.48\textwidth]{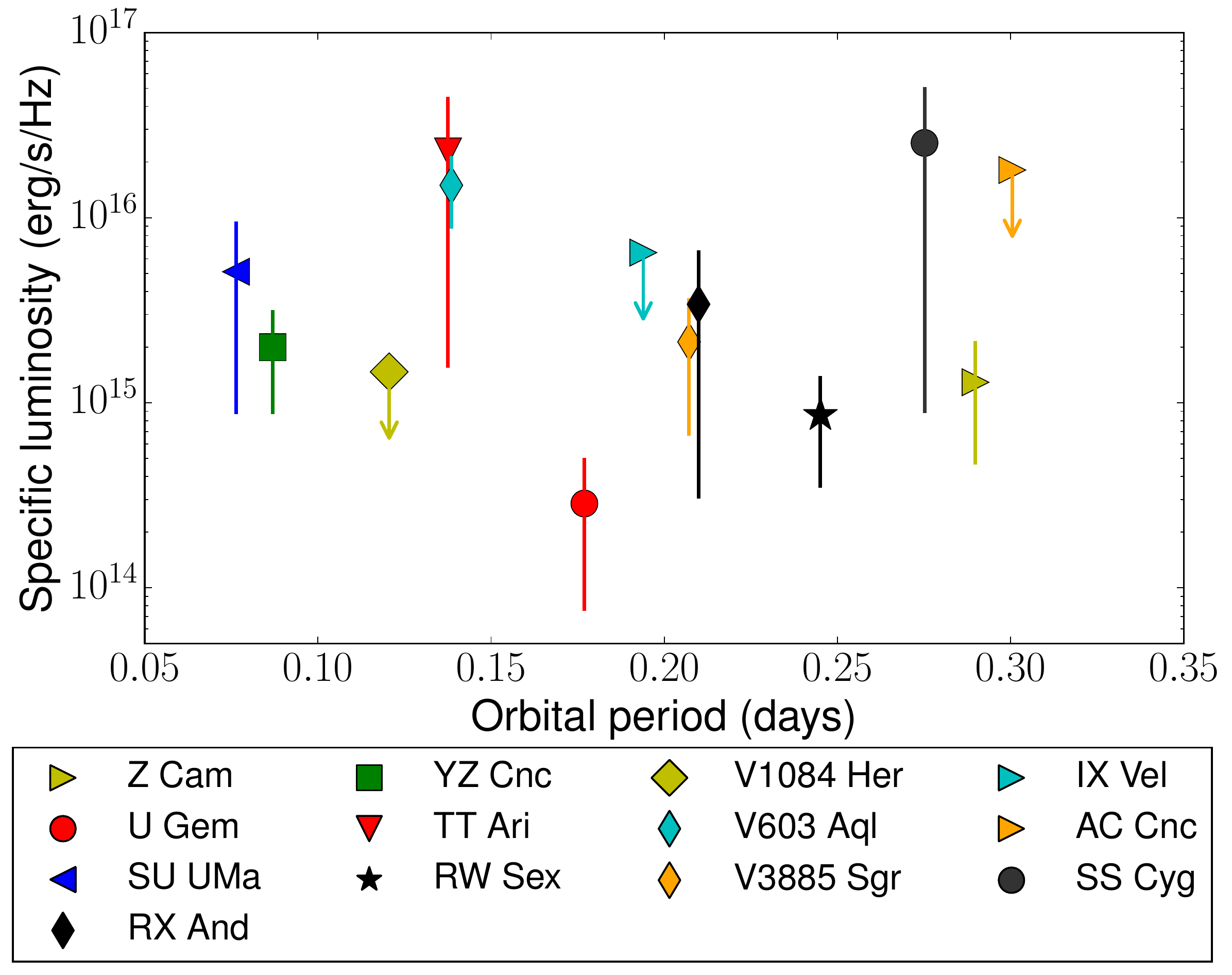}
    \caption{The radio luminosity of non-magnetic CVs is not correlated with orbital period. The error bars indicate the variability range. All high-sensitivity detections of non-magnetic CVs are plotted (this work, \citealt{Koerding2008,Miller-Jones2011,Miller-Jones2013,Coppejans2015,Russell2016}). For a colour version of this figure please see the online material.}
  \label{fig:lum_porb}
\end{figure}

It is not yet clear what parameter determines the radio luminosity of a non-magnetic CV. In the XRBs and AGN, the radio emission in the hard state is directly correlated with the power liberated in the accretion flow \citep{Koerding2006}. \citet{Koerding2008} have suggested that the same is true in CVs.

One proxy for the liberated power is the optical luminosity of the CV. In Figure \ref{fig:radiolum_opticallum}, the radio specific luminosity range is plotted as a function of the quasi-simultaneous optical specific luminosity. Contrary to what we would expect by analogy to the XRBs and AGN, the radio and optical luminosity are not correlated. As the radio emission in non-magnetic CVs can be highly variable however (see Section \ref{sec:variability}, \citealt{Coppejans2015,Russell2016}), sampling effects could be masking an underlying correlation.

Extrapolating from the XRBs, \citet{Koerding2008} estimated that the specific radio luminosity at 10\,GHz ($L_{\rm 10\,GHz}$ in $\rm{erg\,s^{-1}\,Hz^{-1}}$) should be related to the mass accretion rate ($\dot{M}$) by
\begin{equation}
 L_{\rm 10\,GHz}\approx1.5\times10^{24}\frac{\dot{M}}{M_{\odot}\,y^{-1}}\,.
\end{equation}
The mass accretion rate for DN in outburst is estimated to be in the range $10^{-10}$--$10^{-8}\,\rm{M_{\odot}\,y^{-1}}$ (this is model-dependent and is a function of the orbital period, e.g. \citealt{Knigge2011}). This predicts a specific radio luminosity of \power{1}{14}--\power{1}{16}$\rm{erg\,s^{-1}\,Hz^{-1}}$, which is consistent with the observed luminosities.

To establish whether the radio luminosity correlates with the power liberated in the accretion flow, a relative measure of the accretion rate of our systems is necessary. Unfortunately this is not known, as it is difficult to measure in DN.

In our sample we have DN from the three DN-subclasses (U Gem, Z Cam and SU UMa), the novalikes, and observations taken during outburst and superoutburst. As the different classes of CV are predicted to have different secular mass-accretion rates, we looked for a correlation between the radio luminosity and the CV-class. Figure \ref{fig:radiolum_opticallum} does not show this correlation. There were no clear distinctions in the luminosity between the three DN sub-classes, novalikes, or different outburst types, and there was no one class or outburst type that was consistently fainter or brighter. Similarly, plotting the radio luminosity range for each CV as a function of $P_{\rm orb}$ (Figure \ref{fig:lum_porb}) does not show a correlation. The accretion rate in CVs, however, is known to vary significantly on time-scales that are significantly shorter than the secular time-scales \citep[e.g.][]{Livio1994,Smak2004}, due to e.g. variations in the mass transfer rate from the secondary star \citep{King1995,King1996} or nova eruptions \citep{Macdonald1986,Shara1986}. To determine if the radio luminosity is correlated with the accretion flow power, we consequently need measurements of the instantaneous accretion rate for these CVs.  

For completeness, we also compared the radio emission properties in the different CV, and outburst, classes. In each class or outburst type there was at least one CV that was variable on timescales down to minutes (RX And, SU UMa, SS Cyg, TT Ari, V603 Aql). The spectral index was unfortunately not constrained in the DN, so we could not make a comparison on this basis. In all cases, the observed emission was non-thermal, and was consistent with synchrotron or coherent emission. High levels of circular polarization (CP) have been detected in two non-magnetic CVs in different classes. In the novalike TT Ari, the CP fraction peaked at more than 75\% and lasted approximately 10 minutes \citep{Coppejans2015}. In the Z Cam-type DN EM Cyg \citep{Benz1989} the polarization fraction peaked at 81\% in one of the two observations (which were separated by 37 hours). In both cases the radio emission was coherent. The rest of the recent, high-sensitivity observations of non-magnetic CVs have CP fractions $\leq$10-20\%, and linear polarization fractions $\leq$6-15\%. In this sample we do not see a difference in the radio luminosity or properties between the novalikes, different classes of DN, or different classes of outbursts.

Apart from the accretion flow power, other properties such as the WD mass, WD magnetic field strength, inclination angle, or the surrounding medium could affect the radio luminosity. These properties are insufficiently well-constrained in this sample to test this. To determine what property sets the radio luminosity, high cadence observations of a larger sample of CVs with well-determine properties are necessary.   

\section{Summary}\label{sec:conclusion}

Radio emission at a frequency of 8--12\,GHz was detected in all five of the DN (Z Cam, RX And, SU UMa, YZ Cnc and U Gem) that we observed in outburst, which increases the number of radio-detected CVs by a factor of two and proves that DN in outburst are radio emitters with specific luminosities in the range L$_{\rm 10\,GHz}\sim10^{14}$--$10^{16}\,{\rm erg\,s}^{-1}\,{\rm Hz}^{-1}$ at 10\,GHz (luminosities of $10^{24}$--$10^{26}\,{\rm erg\,s}^{-1}$). The emission is variable, and is in excess of the radio emission expected from the CV components (WD, disc and secondary star). Previous radio surveys of DN did not have the sensitivity required to detect these objects, which explains why the detection rate was so low in historical surveys. Combined with the finding that the novalike systems are radio emitters \citep{Coppejans2015}, this indicates that, as a class, non-magnetic CVs are radio emitters.

The emission is consistent with synchrotron or coherent emission, but it is more likely to be synchrotron emission, as the upper-limits on the polarization fraction are on the order of $\sim10$\%, and coherent emission is associated with high levels of polarization. In Z Cam, there is additional support for this, as synchrotron emission at mid-infrared frequencies has been detected by \citet{Harrison2014}.

The DN SS Cyg is a radio synchrotron source in outburst \citep{Koerding2008,Miller-Jones2011,Russell2016} and the emission has been found to originate from a transient jet \citep{Koerding2008,Miller-Jones2011,Russell2016}. We compared the radio emission of our sample of DN to that of SS Cyg, and found that SS Cyg is not unique in the type of radio emission, the luminosity (in the plateau phase), or the variability time-scales (minutes to days). SS Cyg shows rising radio emission on the rise to outburst followed by a radio flare (a key prediction of the jet-launching model, \citealt{Koerding2008}). Although we did not detect a clear equivalent rise in our observations (or a high-amplitude flare), the sampling cadence was insufficient to rule it out. Higher cadence observations over the course of an outburst (and in quiescence) are necessary to establish the radio light-curve template for the DN as a class. High spatial resolution radio observations are needed to resolve extended emission and establish whether CVs, as a class, launch jets.

It is not yet clear what physical property is responsible for setting the radio luminosity. The observed radio luminosity of our sample is consistent with that predicted in \citet{Koerding2008} from outflows based on the accretion-luminosity scaling relations in XRBs and AGN. Since the mass accretion rates in DN are not well determined, we could not test whether the accretion rate is correlated with the radio luminosity, as predicted for an outflow. In this sample there was no correlation between the radio luminosity and optical luminosity (a proxy for the outflow power), the orbital phase or the orbital period. Additionally we did not find any clear distinctions in the radio emission between the different DN classes, novalikes, or normal outbursts and superoutbursts. As our work and that of \citet{Koerding2008}, \citet{Coppejans2015}, \citet{Miller-Jones2011} and \citet{Russell2016} shows, CVs are highly variable and the measured luminosity depends on the sampling time. The low cadence of observations in our sample (relative to either the one well-sampled DN, SS Cyg, or the evolving physical parameters,) could consequently mask any underlying correlations.

\section*{Acknowledgements}

This work is part of the research programme NWO VIDI grant No. 2013/15390/EW, which is financed by the Nederlandse Organisatie voor Wetenschappelijk Onderzoek (NWO). DLC gratefully acknowledges funding from the Erasmus Mundus Programme SAPIENT. JCAMJ is the recipient of an Australian Research Council Future Fellowship (FT140101082). GRS acknowledges support from an NSERC Discovery Grant. CK is grateful for support from the Science and Technology Facilities Council under grant ST/M001326/1 and from the Leverhulme Trust for the award of a Research Fellowship. PAW thanks UCT and the NRF for financial support.

The authors are grateful to the observers from the American Association of Variable Star Observers (AAVSO), whose observations and notifications were crucial to this research. We would like to especially thank the following observers for their significant contributions: Umair Asim, Douglas Barrett, Paul Benni, Richard Campbell, Shawn Dvorak, Tonis Eenmae, Steve Gagnon, Josch Hambsch, Dave Hinzel, Laszlo Kocsmaros, Mike Linnolt, Walter MacDonald, Chris Maloney, Vance Petriew, Gary Poyner, Anthony Rodda, Paolo Ruscitti, Rod Stubbings, Glenn Thurman, John Toone, Brad Vietje and Brad Walter.

The National Radio Astronomy Observatory is a facility of the National Science Foundation operated under cooperative agreement by Associated Universities, Inc. This research has made use of NASA's Astrophysics Data System Bibliographic Services, as well as the SIMBAD data base, operated at CDS, Strasbourg, France (Wenger et al. 2000).



\bibliographystyle{mnras}
\bibliography{master_bibliography.bib} 


\appendix

\section{Higher-time resolution light curves}

\begin{table*}
  \centering
  \begin{minipage}{140mm}
    \caption{Higher time resolution}
    \begin{tabular}{|l|l|l|l|r|l|}
    \hline
    Object & Obs. & MJD & Orbital phase$^a$ & Integration & Flux Density\\
    &  & (mid-exposure) & (mid-exposure) & time (s) & ($\mu$Jy)\\
    \hline
    Z Cam & 1 & 56986.12344 & 0.66 $\pm$ 0.02$^b$ & 2053 & <18.3\\
    & 1 & 56986.14720 & 0.74 $\pm$ 0.02$^b$ & 2053 & 28.8 $\pm$ 5.1\\
    & 1 & 56986.17096 & 0.82 $\pm$ 0.02$^b$ & 2053 & 27.3 $\pm$ 5.9\\
    Z Cam & 2 & 56987.17910 & 0.30 $\pm$ 0.02$^b$ & 1254 & 34.2 $\pm$ 7.2\\
    & 2 & 56987.19361 & 0.35 $\pm$ 0.02$^b$ & 1254 & 46.7 $\pm$ 7.0\\
    Z Cam & 3 & 56988.15027 & 0.65 $\pm$ 0.02$^b$ & 737 & 29.3 $\pm$ 8.0\\
    & 3 & 56988.15880 & 0.68 $\pm$ 0.02$^b$ & 737 & 29.4 $\pm$ 8.8\\    
    & 3 & 56988.16733 & 0.71 $\pm$ 0.02$^b$ & 737 & 41.3 $\pm$ 8.5\\
    RX And & 1 & 56969.42185 & 0.19 $\pm$ 0.03$^c$ & 3095 & 14.3 $\pm$ 4.3\\
    & 1 & 56969.45768 & 0.36 $\pm$ 0.03$^c$ & 3094 & 13.5 $\pm$ 4.5\\
    RX And & 2 & 56970.36307 & 0.68 $\pm$ 0.03$^c$ & 650 & <27.3\\
    & 2 & 56970.37059 & 0.71 $\pm$ 0.03$^c$ & 650 & <26.7\\
    & 2 & 56970.37811 & 0.75 $\pm$ 0.03$^c$ & 650 & <27.3\\
    & 2 & 56970.38375 & 0.78 $\pm$ 0.03$^c$ & 325 & <40.2\\
    & 2 & 56970.38657 & 0.79 $\pm$ 0.03$^c$ & 163 & 77.0 $\pm$ 20.0\\
    & 2 & 56970.38846 & 0.80 $\pm$ 0.03$^c$ & 163 & 95.0 $\pm$ 15.0\\
    SU UMa & 1 & 57027.16243 & 0.88 $\pm$ 0.19$^d$ & 1986 & 44.6 $\pm$ 6.5\\
    & 1 & 57027.18542 & 0.18 $\pm$ 0.19$^d$ & 1986 & 30.3 $\pm$ 6.2\\
    & 1 & 57027.20840 & 0.49 $\pm$ 0.19$^d$ & 1986 & 34.8 $\pm$ 6.2\\
    SU UMa & 2 & 57028.07291 & 0.81 $\pm$ 0.19$^d$ & 797 & 83.1 $\pm$ 8.7\\
    & 2 & 57028.08213 & 0.92 $\pm$ 0.19$^d$ & 797 & 45.1 $\pm$ 8.6\\
    & 2 & 57028.09135 & 0.05 $\pm$ 0.19$^d$ & 797 & 33.4 $\pm$ 9.0\\
    SU UMa & 3 & 57029.43929 & 0.70 $\pm$ 0.19$^d$ & 1210 & 21.9 $\pm$ 6.9\\
    & 3 & 57029.45330 & 0.88 $\pm$ 0.19$^d$ & 1210 & <21.9\\
    YZ Cnc & 1 & 56984.30181 & 0$^{e,f}$ & 3024 & <16.2\\
    & 1 & 56984.33681 & 0.40265 $\pm$ 0.00003$^{e,f}$ & 3024 & 20.6 $\pm$ 5.5\\
    U Gem & 1 & 57075.34642 & 0.908 $\pm$ 0.002$^g$ & 3000 & <12.9\\
    & 1 & 57075.38115 & 0.104 $\pm$ 0.002$^g$ & 3000 & 12.6 $\pm$ 4.1\\
    U Gem & 2 & 57080.15979 & 0.117 $\pm$ 0.002$^g$ & 1110 & 27.0 $\pm$ 7.7\\
    & 2 & 57080.17264 & 0.189 $\pm$ 0.002$^g$ & 1110 & <23.7\\
    \hline    
    \multicolumn{6}{p{14cm}}{\footnotesize{Notes: When the SNR of an observation (in Table \ref{tbl:results}) was sufficiently high, we split it in time to probe the variability on shorter time-scales. Any observation not listed in the table did not have a sufficiently high SNR. $^a$The orbital phase is absolute unless indicated otherwise, and phase zero corresponds to inferior conjunction. Note that the uncertainty on the orbital phase is calculated at mid-exposure, it does not taken the integration time into account. The ephemerides used are from $^b$\citet{Thorstensen1995}, $^c$\citet{Kaitchuck1989}, $^d$HJD = 2450247.986(3)+0.07637533(13) (updated ephemeris from \citet{Thorstensen1986}, private communication with John Thorstensen), $^e$\citet{vanParadijs1994} and $^g$\citet{Marsh1990} updated in \citet{Echevarria2007}. $^f$Phase relative to JD 2456984.801806, as the error on the absolute phase exceeds 0.5 and a recent ephemeris is not available.}}\\
    \end{tabular}
    \label{tbl:lightcurves}
  \end{minipage}
\end{table*}

\bsp	
\label{lastpage}
\end{document}